\documentclass[paper,nohyper]{JHEP3} 

\usepackage{epsfig,multicol}
\usepackage{cite}
\usepackage{amsmath}

\newcommand\fverb{\setbox\pippobox=\hbox\bgroup\verb}
\newcommand\fverbdo{\egroup\medskip\noindent%
            \fbox{\unhbox\pippobox}\ }
\newcommand\fverbit{\egroup\item[\fbox{\unhbox\pippobox}]}
\newbox\pippobox
\newcommand{\as}{\overline{\alpha}_s}

\newcommand{\mz}     {\ensuremath{M_{\rm Z}}}
\newcommand{\ymax}     {\ensuremath{y_{\rm max}}}

\newcommand{\bm}[1]  {\mbox{\boldmath $#1$}}
\newcommand{\epem}     {\ensuremath{{\rm e}^+{\rm e}^-}}
\newcommand{\bt}     {\ensuremath{B_T}}
\newcommand{\bn}     {\ensuremath{B_N}}
\newcommand{\bw}     {\ensuremath{B_W}}

\newcommand{\oaa}    {\ensuremath{\mathcal{O}(\alpha_s^2)}}
\newcommand{\order}[1]{\ensuremath{\mathcal{O}(#1)}}
\newcommand{\alphab} {{\overline \alpha_s}}
\newcommand{\rz}     {\ensuremath{R_{\rm Z}}}

\newcommand{\xmu}   {\ensuremath{x_\mu}}

\newcommand\Govrln{{\widehat G}}

\newcommand\gae{{\gamma_{{E}}}}
\newcommand\asb{{\bar \alpha}_{{\textsc{s}}}}
\def\half{\mbox{\small $\frac{1}{2}$}}

\def\Govrln{{\overline G}}

\def\as{\alpha_{s}}
\def\gae{{\gamma_{\textsc{e}}}}
\def\asb{{\bar \alpha}_{{\textsc{s}}}}

\title{Theoretical uncertainties on $\boldsymbol{\alpha_s}$ from event-shape variables in $\boldsymbol{\epem}$ annihilations}

\author{Roger~W.~L.~Jones \\ Lancaster University \\ Lancaster, LA1 4Yb, UK
 \\ Email: \email{Roger.Jones@cern.ch}}
\author{Matthew~Ford \\ Cavendish Laboratory, Madingley Road \\ Cambridge, CB3 0HE, UK
 \\ Email: \email{Matthew.Ford@cern.ch}}
\author{Gavin~P.~Salam \\ LPTHE, Universit\'es Paris VI and VII, and CNRS
  UMR 7589\\
Paris 75005, France \\ Email: \email{salam@lpthe.jussieu.fr}}
\author{Hasko~Stenzel \\ II. Physikalisches Institut, JLU Giessen \\
Heinrich-Buff Ring 16, D-35392 Giessen, Germany \\ Email: \email{Hasko.Stenzel@exp2.physik.uni-giessen.de}}
\author{Daniel~Wicke \\ Bergische Universit\"at \\
Gau{\ss}str. 20, D-42097 Wuppertal
 \\ Email: \email{Daniel.Wicke@physik.uni-wuppertal.de}}

\abstract{The precision of measurements of the strong coupling
  constant using event-shape variables in $\epem$ annihilations is
  limited by theoretical systematic uncertainties. The uncertainties
  are related to missing higher orders in the perturbative predictions
  for the event-shape distributions. A new method is presented for the
  assessment of theoretical uncertainties in $\alpha_s$.  This method
  evaluates the systematic uncertainty of the parameter $\alpha_s$
  from the uncertainty of the prediction for the distributions from which
  it is extracted. The perturbative uncertainties are calculated
  on a purely theoretical basis, without accessing measured distributions.
  The method is therefore especially suited for an unbiased combination of
  results from different observables or experiments.
  It is universal and can be applied to other
  processes like jet production in deep-inelastic $ep$ scattering or in
  hadron collisions.  }
\received{November 30th, 2003}        
\revised{December 15, 2003}
\accepted{December 20, 2003}        

\preprint{}  

\begin{document}


\section{Introduction}
Studies of Quantum Chromodynamics in $\epem$ annihilations have been
carried out over more than 30 years at increasing centre-of-mass
energies. Event-shape variables have proven to be key observables in
both annihilation and deep-inelastic scattering processes.  The
understanding of perturbative and non-perturbative aspects of QCD has
grown with the study of event shapes. While event shapes were first
introduced to characterise global properties, it was soon realised
that their distributions are sensitive to the strong coupling constant
$\alpha_s$. The perturbative prediction for a generic
infrared-collinear (IRC) safe event-shape variable $y$ can be computed
to second order in $\as$,
\begin{equation}
\frac{1}{\sigma}\frac{d\sigma}{dy} =  A(y) \frac{\alpha_s}{2\pi} +B(y) \left(\frac{\alpha_s}{2\pi}\right)^2 \; ,
\end{equation}
with coefficient functions $A$ and $B$ obtained from integration of
the ERT\cite{ERT} matrix elements.  Using this type of prediction,
first determinations of $\alpha_s$ were performed at the PEP and
PETRA colliders. In the era of the LEP experiments the calculations
for certain classes of variables were improved by resumming leading
and next-to-leading logarithmic terms (NLL). These calculations,
matched to fixed-order expressions, enlarged the kinematic range of
applicability for $\alpha_s$ extractions and reduced the
systematic theoretical uncertainty. Recently, event-shape variables
have also been used extensively to study non-perturbative power law
corrections. It has been shown that hadronisation corrections, scaling
with inverse powers of the momentum transfer $Q$ can be modelled with
one or a few non-perturbative parameters \cite{milano}. These
parameters can in turn be related to moments of an effective coupling
at low scales, and have been extracted from the data \cite{leppowcor}.

At present measurements of $\alpha_s$ using event-shape variables and
$\oaa + $NLL resummed calculations are available at centre-of-mass
energies between 22 GeV and 206 GeV. For illustration a selection of
these measurements is shown in Fig.~\ref{fig:asrun}.
\FIGURE{\epsfig{file=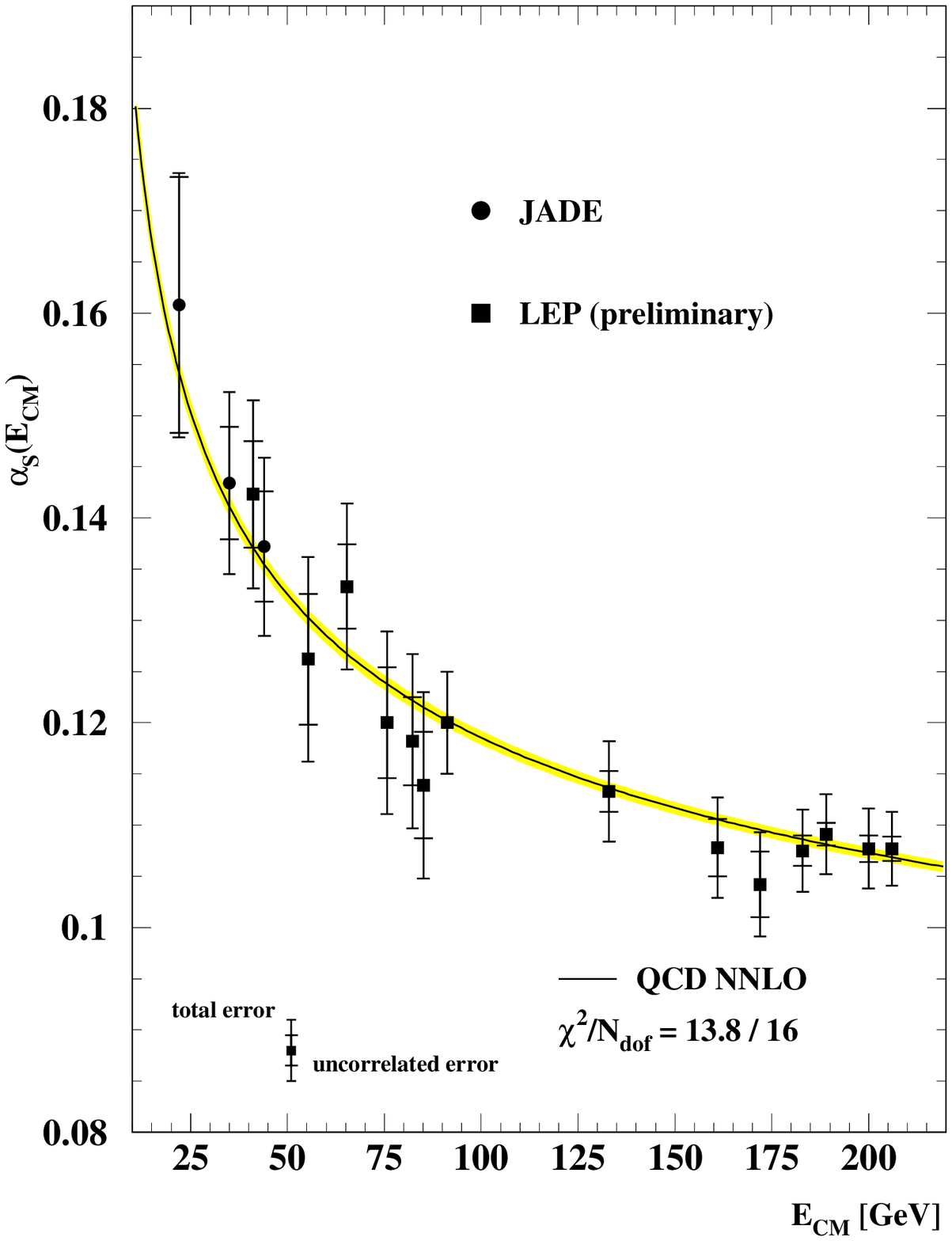,height=120mm}
\caption{Measurements of $\alpha_s$ from event-shape variables as function
  of centre-of-mass energy.}\label{fig:asrun}}

New measurements re-analysing data from the JADE collaboration
\cite{jade} were performed using a technique similar to that applied
for a combined analysis by the LEP collaborations \cite{qcdwg}. In
particular, the same theoretical framework was used and at most of the
energies a similar set of variables was combined. At almost all energies
perturbative theoretical uncertainties dominate other systematic
uncertainties, e.g. at $Q=M_{\rm Z}$ the perturbative uncertainty is $5\%$
and the quadratic sum of statistical, experimental and
hadronisation uncertainties is $1.3\%$.

It is the objective of this note to examine ways of estimating the
theoretical uncertainties and to propose a comprehensive and
consistent method to combine different uncertainty estimates. In
addition, the proposed method should allow the calculation of the
uncertainty associated with a given measurement without using the
 data from which $\alpha_s$ is extracted. This is notably
important for global combinations of various measurements, where a
coherent procedure must be applied.

The origin of
perturbative theoretical uncertainties is the truncation of the perturbative
series
at a given order of the coupling constant,
whereby the truncated prediction acquires a
dependence on the renormalisation scale.


To illustrate this point, a fully inclusive
observable, the total hadronic cross section $\sigma(\epem
\rightarrow {\rm q\bar q})$, is considered as an example for which third
order $\mathcal{O}(\alpha_s^3)$
calculations are available.  The ratio $R_{\rm Z}$ of hadronic to
leptonic widths at the Z boson resonance reads (with renormalisation
scale $\mu = Q$) \cite{leprz}
\begin{equation}
\label{eq:rz}
\rz = \frac{\Gamma({\rm Z^0 \rightarrow hadrons})}{\Gamma({\rm Z^0 \rightarrow leptons})} =
R_{\rm EW} \sum_{n=0}^3 c_n \left(\frac{\alpha_s}{\pi}\right)^n
\end{equation}
with
\begin{equation}
\label{eq:rzcoeff}
R_{\rm EW} = 19.934 \; , \; \; c_0=1 \; , \; \; c_1=1.045 \;  , \; \;  c_2 = 0.94 \; , \; \; c_3=-15\,,
\end{equation}
%
%
%
One estimate of the theoretical uncertainty on $R_{\rm
  Z}$ is based on the size of the last term in the series, taking the
difference between the full next-to-next-to-leading
order (NNLO) prediction and the next-to-leading order (NLO) truncation. Taking the
preliminary combined LEP measurement $R_{\rm Z}= 20.767 \pm 0.025$
\cite{LEPEW} the NNLO result is $\alpha_s(M_{\rm Z})=0.1240 \pm
0.0037(exp)$ and the central value shifts down to
$\alpha_s^{NLO}(M_{\rm Z})=0.1214$ when the NLO prediction is used.
Hence, the theoretical uncertainty estimated by the NNLO to NLO
difference is $\Delta\alpha_s = \pm 0.0026$. The `true' error would be
the difference between the NNLO result and the complete theory. The
NNLO to NLO difference can only be a good estimate of that if the
convergence of the perturbation series is fast enough, {\it i.e.} if the
size of the expansion coefficients remains of order unity. In the case
of $\rz$, however, the size of the third order term amounts to 60 $\%$
of the second order term size.

Of course the size of the last known term of the series is not the
only way of estimating the uncertainty on the prediction. Another
method originates from the fact that dimensional
regularisation, used to define formally divergent loop corrections
prior to $\overline{\mbox{\textsc{ms}}}$ renormalisation, requires
the introduction of a renormalisation scale $\mu$ at which
the coupling, $\as(\mu^2)$, is defined. While the choice of scale is arbitrary,
higher-order perturbative corrections contain terms proportional to
powers of $\ln \mu^2/Q^2$, which order-by-order compensate for the
scale dependence of the coupling. Only in the presence of all orders
this compensation is complete. For a calculation to order $\as^n$
there is residual scale-dependence of order $\as^{n+1}$; since it
would be cancelled by higher order terms, its size can be taken as
indicative of the magnitude of missing higher-order contributions.


The scale dependence of Eq.~(\ref{eq:rz}) is given by the dependence
of the NLO and NNLO coefficients on the scale parameter
$x_\mu=\frac{\mu}{Q}$ as follows
\begin{eqnarray}
c_2(x_\mu) & = & c_2^0 + \beta_0 c_1 \ln x_\mu^2 \\
c_3(x_\mu) & = & c_3^0 - \left(2 \beta_0 c_2^0 + c_1 \beta_1\right) \ln x_\mu^2
+ c_1\beta_0^2\ln^2x_\mu^2 \; ,
\end{eqnarray}
where the coefficients $c_i^0$ are those of Eq.~(\ref{eq:rzcoeff}) and
the first three coefficients of the $\beta$ function are
\begin{equation}
\beta_0 = \frac{33-2 \, n_f}{12} \; , \; \; \beta_1 = \frac{153-19 \, n_f}{24} \; ,
\; \; \beta_2  =  \frac{1}{64}\left[\frac{2857}{2} - \frac{5033}{18}\; n_f + \frac{325}{54} \; n_f^2 \right] \; .
\end{equation}
The number of active flavours $n_f$ is taken to be five at $Q=M_{\rm
  Z}$. The renormalisation scale dependence of the
coupling constant can be parameterised in the modified minimal
subtraction scheme at 3-loop level \cite{running} as a function of the
scale $\Lambda \equiv \Lambda^{(nf)}_{\rm\overline MS} $
\begin{eqnarray}
\label{running_formula}
\alpha_s(\mu) & = & \frac{\pi}{\beta_0\ln(\mu^2/\Lambda^2)}\left[
1-\frac{\beta_1}{\beta_0^2}\frac{\ln\left[\ln(\mu^2/\Lambda^2)\right]}
{\ln(\mu^2/\Lambda^2)}
+\frac{1}{\beta_0^2\ln^2(\mu^2/\Lambda^2)}\right. \nonumber \\*[0.2 cm]
& & \left. \times \left( \frac{\beta_1^2}{\beta_0^2}\left\{\ln^2(\mu^2/\Lambda^2)-\ln\left[\ln(\mu^2/\Lambda^2)\right]
-1 \right\} +
\frac{\beta_2}{\beta_0}\right)\right] \; \; .
\end{eqnarray}
The dependence of the measurement of $\alpha_s(\mz)$ using $\rz$
is illustrated in Fig.~\ref{fig:rzfit}.
\FIGURE[t]{\epsfig{file=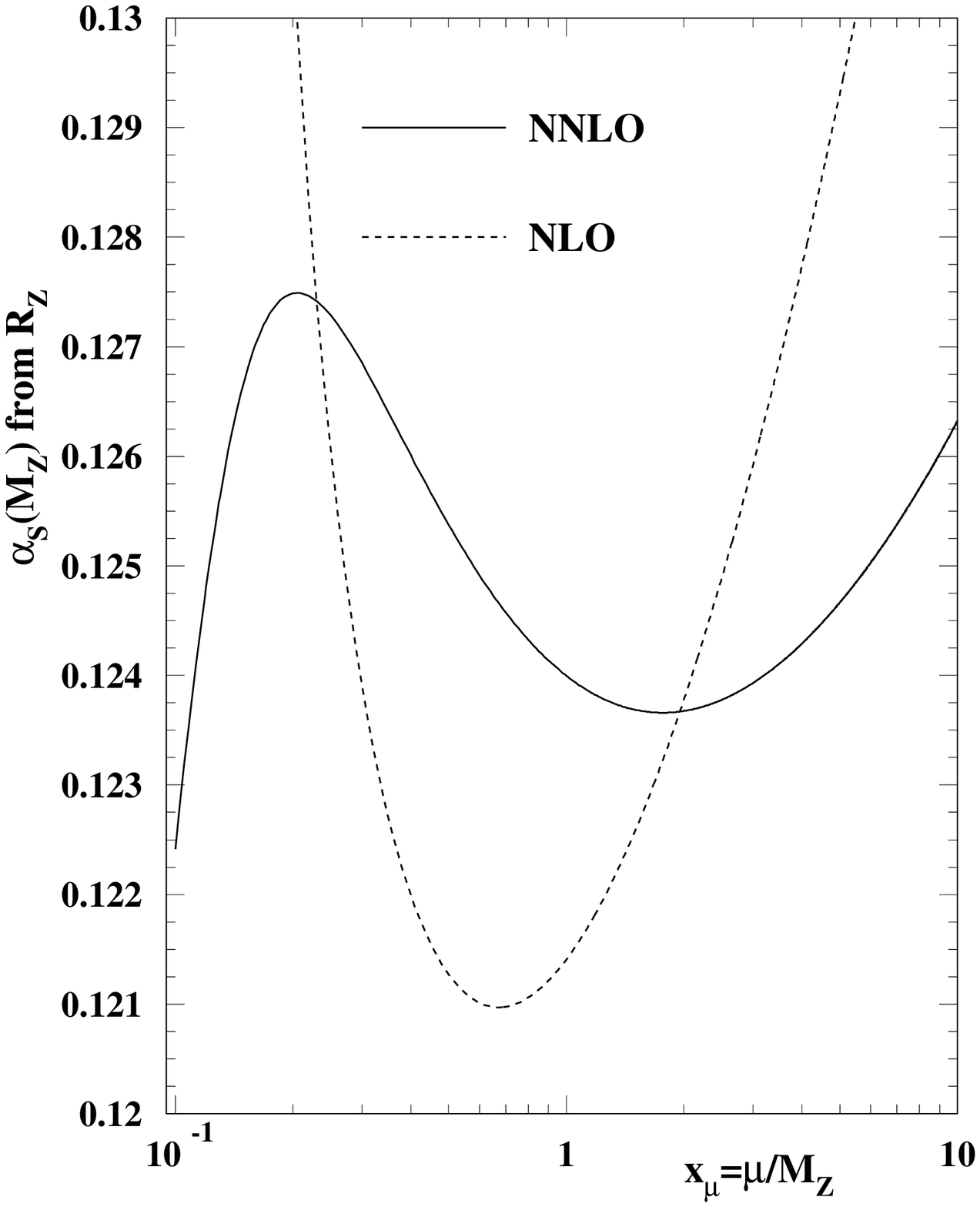,height=120mm}
\caption{The dependence on the
  renormalisation scale parameter of $\alpha_s(M_{\rm Z})$ using $\rz$
  at NNLO (solid line) and at NLO (dashed line).}\label{fig:rzfit}}
The case of $\rz$ nicely demonstrates the evolution of the scale
dependence from NLO to NNLO. The structure of the perturbative
series indicates that a natural value for the scale $\mu$ is of
order $\mz$, but its exact value is undetermined.



In $\epem$ annihilation it has become customary to take a central
value of $x_\mu =1$ and to estimate an uncertainty in the prediction
by varying $x_\mu$ between $\frac{1}{2}$ and 2. (In contrast, in global
analyses of parton distribution functions it is sometimes $x_\mu^2$
that is varied in that range, e.g.~\cite{ZEUS}.)  A variation of
$x_\mu$ from
$\frac{1}{2}$ to 2 applied to $\rz$ yields a theoretical uncertainty
for $\alpha_s$ of $\pm 0.0020$, slightly lower than, but of the same
general size as that obtained from the NNLO to NLO difference. This is
despite the fact that the uncertainty from the $x_\mu$ variation is
formally higher order in $\as$ than that from the NNLO to NLO
difference.

In contrast to $\rz$, event-shape distributions, which shall be
studied in this paper, are known only to next-to-leading accuracy. In
principle here too the last term of the series could be used to
estimate the uncertainty.  However, this is often seen as unduly
conservative (it would give an uncertainty of order $\as^2$, whereas
missing terms are of order $\as^3$) and the standard approach for NLO
fixed-order calculations is to use just scale variations to estimate
uncertainties.

Also in contrast to the $\rz$ case, over a significant part of the
phase-space the event-shape distributions have \emph{divergent} (or very
poorly convergent)
perturbative expansions, due to logarithmic enhancements of the
perturbative coefficients. As a result, it is necessary to supplement
fixed-order results with resummed calculations
\cite{tmresum,CTTW,yresum,newbresum,cresum,gaviny3}, leading to a
series with two expansion parameters --- the coupling and the
logarithm of the observable. Estimating uncertainties on the result is
considerably more involved than for a traditional fixed-order
expansion. For example, in the same way that the presence of a formally
arbitrary scale $x_\mu$ for the coupling is associated with an
uncertainty, a new formally arbitrary scale $x_L$ appears in the
definition of the logarithm and it too is associated with an
uncertainty on the perturbative prediction. Further uncertainties
arise from freedom in the way resummed and
fixed-order predictions are combined (`matched').

A number of these possible sources of uncertainty have come to light
only recently, partly in the context of DIS event-shape resummations
\cite{DISXL}. There is therefore a need for a comprehensive up to date
study of the whole range of such uncertainties for $\epem$ event
shapes, as well as an understanding of how to combine them. That is
the purpose of the present article.


The definitions of the event-shape variables are given in Section
\ref{sec:var}, in Section \ref{sec:th} the theoretical predictions are
summarised, in Section \ref{sec:error} the methods for estimating and
combining theoretical uncertainties are presented, in Section
\ref{sec:results} results obtained with the new method are discussed
and the recommendation for an uncertainty strategy is summarised in
Section \ref{sec:conc}.

\section{Definition of the event-shape variables}
\label{sec:var} Event-shape variables are infrared and collinear
safe observables that characterise global properties of hadronic
events. They are experimentally measured with the momenta of
charged and/or neutral particles. The variables are normalised so
as to be dimensionless and independent of the production
cross-section. One end of the spectrum is given by the two-jet
limit, the other by a kinematic limit $\ymax$, which in some cases
corresponds to either spherical or symmetric event configurations.
The following variables are studied here.
\begin{description}
\item[Thrust \bm{T} \cite{def_t}:]
  the thrust axis unit vector $\vec{n}_T$
  maximises the following quantity:
  \begin{displaymath}
    T= \max_{\vec{n}_T}\left(\frac{\sum_i|\vec{p}_i\cdot\vec{n}_T|}
      {\sum_i| \vec{p}_i|}\right) \; , \; \; \tau=1-T \; ,
  \end{displaymath}
  where the thrust sum extends over all particles $i$ in the event. It is convenient to
define $\tau=1-T$ for the resummed expressions.
\item[Heavy Jet Mass \bm{\rho} \cite{def_m}:] a plane perpendicular to
  $\vec{n}_T$ divides the event into two hemispheres, $H_1$ and $H_2$,
  from which one obtains the corresponding normalised hemisphere
  (squared) invariant masses:
  \begin{displaymath}
  M_i^2 = \left(\sum_{k \in H_i} p_k \right)^2 \;\; , \; i=1,2 \;\; .
  \end{displaymath}
  The larger of the two hemisphere masses is called the heavy jet mass :
  \begin{displaymath}
  \rho \equiv M_H^2 = \frac{1}{E_{vis}^2}\max(M_1^2,M_2^2) \;\; ,
  \end{displaymath}
  where $E_{vis}$ is the total visible energy in the event.
\item[Wide Jet Broadening \bm{\bw} \cite{def_b}:]
  a measure of the broadening of particles in transverse momentum
  with respect to the thrust axis can be calculated for each hemisphere
  $H_i$ using the relation:
  \begin{displaymath}
    B_i= \frac{\sum_{k\in H_i}|\vec{p}_k\times \vec{n}_T|}
                    {2\sum_j|\vec{p}_j|}  \; \; , \; i=1,2 \; \;\; \; ,
  \end{displaymath}
  where $j$ runs over all of the particles in the event.
  The wide jet broadening is the larger of the two hemisphere broadenings:
  \begin{displaymath}
      \bw = \max (B_1,B_2) \; \; .
  \end{displaymath}
  Equivalently, the narrow jet broadening  \bm{\bn} is defined as $\bn = \min (B_1,B_2)$.
\item[Total Jet Broadening \bm{\bt}:]
  the total jet broadening is the sum of the wide jet broadening and the narrow jet broadening:
  \begin{displaymath}
      \bt = \bw + \bn \; \; .
  \end{displaymath}
\item[C-parameter \bm{C} \cite{def_c}:]
  the C-parameter is derived from the eigenvalues of the linearised \\
momentum tensor $\Theta^{\alpha\beta}$ :
  \begin{displaymath}
    \Theta^{\alpha\beta}= \frac{1}{\sum_i|\vec{p}_i|}
    \sum_i \frac{p_{i}^\alpha p_{i}^\beta}
                               {|\vec{p}_i|}\;\;,
                           \;\alpha,\beta= 1,2,3\;\;.
  \end{displaymath}
  The three eigenvalues $\lambda_j$ of this tensor define $C$ with:
  \begin{displaymath}
    C= 3\cdot(\lambda_1\lambda_2+\lambda_2\lambda_3+\lambda_3\lambda_1) \;.
  \end{displaymath}
\item[Durham jet resolution parameter \bm{y_3} \cite{Durham}:] the
  Durham clustering algorithm for jet rates is taken as follows.  For
  each pair of particles $i$ and $j$ in an event the metric $y_{ij}$ is computed:
  \begin{displaymath}
    y_{ij} =
    \frac {2 \, \mbox{min}(E_i^2, E_j^2) (1 - \cos \theta_{ij})}
    {E_{vis}^2} \;\;\; \mbox{\rm (Durham algorithm)} \; \; .
  \end{displaymath}
  The pair of particles with the smallest value of $y_{ij}$ is
  replaced by a cluster.  The four-momentum of the cluster is taken to
  be the sum of the four momenta of particles $i$ and $j$, $p^{\mu} =
  p_i^{\mu} + p_j^{\mu}$ (`E' recombination scheme).  The clustering
  procedure is repeated until all $y_{ij}$ values exceed a given
  threshold $y_\mathrm{cut}$.  The number of clusters remaining at
  this point is defined to be the number of jets.

  The $y_3\equiv y_{23}$ jet resolution parameter is the threshold
  value of $y_{\mathrm{cut}}$ below which an event is classified as
  having two jets and above which it has three jets. The distribution of
  $y_3$ falls steeply, with only a very narrow peak at small $y_3$.
  In order to better examine the region of small $y_3$, the
  logarithmic form $-\ln y_3$ is usually analysed.
\end{description}

\section{Theoretical predictions}
\label{sec:th}

In this section the ingredients of the theoretical
calculations as well as methods for gauging the uncertainties are outlined. At the end
of the section directions in which there is partial
theoretical progress in improving the accuracies of the theoretical
calculations are mentioned.

\subsection{Theoretical ingredients}

\paragraph{Fixed order calculation}

To second order in $\alpha_s$, the distribution of
a generic event-shape variable $y$ ($y$=$\tau$, $\rho$, $\bt$, $\bw$, $C$ or $y_3$) is given by:
 \begin{eqnarray}\label{fixed}
\frac{1}{\sigma_{tot}}\; \frac{d\sigma(y)}{dy}&=&
\alphab(\mu^{2})A(y)+\left(\alphab(\mu^{2})\right)^{2} \left[
A(y)\, 2 \beta_0 \ln \xmu^2 +B(y)\right] \; \; ,
\end{eqnarray}
\begin{equation}
\mbox{where} \; \;
\alphab=\frac{\alpha_s}{2\pi}\; , \;\; \xmu=\frac{\mu}{Q} \; , \; \;   \mu = \mbox{renormalisation scale}.
\end{equation}
The coefficient functions $A$ and $B$ are obtained from
integration of the ERT matrix elements, using for instance the
integration program EVENT2 \cite{EVENT2}. Consider the cumulative
cross section:
\begin{equation}
R(y,\alpha_s) \equiv \frac{1}{\sigma_{tot}}\int_{0}^{y}\frac{d\sigma(x,\alpha_s)}{dx}dx \; \; ,
\end{equation}
which may be cast into the second-order form
\begin{equation}
R^{{\cal O}(\alpha_{s}^{2})}(y,\alpha_s) = 1 + {\cal A}(y)\alphab +
\left[{\cal A}(y) \, 2  \beta_0 \ln x_\mu^2 + {\cal B}(y) \right]\alphab^2 \;  ,
\end{equation}
where ${\cal A}$ and ${\cal B}$ are integrated forms of $A$ and $B$, and the explicit scale dependence of $\alpha_s$ has
been dropped for clarity.

\paragraph{Resummed calculations}

For small values of $y$, the fixed-order expansion, Eq.~\eqref{fixed}
fails to converge, because the fixed-order coefficients are enhanced
by powers of $\ln 1/y$, ${\cal A}(y) \sim \ln^2 y$, ${\cal B}(y) \sim
\ln^4 y$. To obtain reliable predictions in the region of $y \ll 1$ it
is necessary to \emph{resum} entire sets of logarithmic terms at all
orders in $\alpha_s$. Certain event shapes and jet rates have the property that
double logarithms exponentiate allowing one to write
\begin{equation}\label{eq:Resummed35}
  R(y,\as) = \left(1 + C_1 \alphab + C_2 \alphab + \ldots\right)
  e^{ L g_1(\as
    L) + g_2 (\as L) + \as g_3(\as L) + \ldots} + \order{\as y} \,,
\end{equation}
where $L = \ln y_0/y$, with $y_0=1$ for $y=\tau$, $\rho$, $y_3$, $B_T$,
$B_W$ and $y_0=6$ for $C$-parameter. The function $g_1(\as L)$ resums
leading logarithms (LL), while $g_2(\as L)$ resums next-to-leading
logarithms (NLL), etc. Such a resummation scheme allows to make
reliable predictions down to the region $\as L \sim1$.\footnote{It is
  to be noted that two different logarithmic classifications schemes
  are in use in the literature. The convention adopted here classifies
  the logarithms in $\ln R$ and stems from \cite{CTTW}. In certain other
  contexts (e.g.\ \cite{Durham}) it is logarithms in $R$ itself that
  are classified, so that LL means terms $\as^n L^{2n}$ in $R$, NLL
  gives terms $\as^n L^{2n-1}$, etc. Such a resummation scheme is
  valid in a more restricted range of $L$, $\as L^2 \lesssim 1$.} %
The $g_n(\as L)$ functions have expansions
\begin{equation}
  g_n(\as L) = \sum_{i=1}^\infty G_{i,i+2-n} \alphab^i L^{i+2-n} \,.
\end{equation}
Expressions for the LL and NLL functions $g_1$ and $g_2$ have been
derived for a range of observables
\cite{tmresum,CTTW,yresum,newbresum,cresum,gaviny3}. Additionally the
$C_1$ coefficients are known analytically, and the further subleading
terms $G_{21}$ and $C_2$ are known numerically
\cite{tmresum,CTTW,yresum,newbresum,cresum,gaviny3,gardi}. Thus the
current state of knowledge for resummed results can be written
\begin{equation}
\label{eq:RNLL}
R_{NLL}(y,\alpha_s)  =  \left(1 + C_1 \alphab + C_2 \alphab^2\right)
\exp ( L g_1(\alphab L) + g_2(\alphab L) + G_{21} L\alphab^2 ) \, .
\end{equation}
The full set of coefficients to $\order{\alpha_s^2}$ is given in
Table~\ref{tab:coeffs}.
\TABULAR[h]{|c|c|c|c|c|c|c|}{ \hline
         &   T    & $\rho$ &  C   & $ B_W$ & $B_T$ & $-\ln y_3$ \\ \hline
$C_1$    &  1.053 & 1.053     & 5.44   & 1.826  & 1.826  & $-6.685$ \\ \hline
$C_2$    &  34    & 40        & 76.5   & 116.3  & 92     & 18.2\\ \hline
$G_{11}$ &  4     & 4         & 4      & 8      & 8      & 4 \\ \hline
$G_{12}$ &  $-2.67$ & $-2.67$     & $-2.67$  & $-5.33$  & $-5.33$  & $-4/3$ \\ \hline
$G_{21}$ &  22    & 36        & 63.4   & 73.8   & 78.5   & $-7.2$ \\ \hline
$G_{22}$ & $-24.94$ & $-13.24$    & $-24.94$ & $-15.09$ & $-61.88$ & 0.868\\ \hline
$G_{23}$ & $-10.22$ & $-10.22$    & $-10.22$ & $-27.26$ & $-27.26$ & $-3.407$\\ \hline
}{\label{tab:coeffs}Numerical values of the resummation coefficients,
  to $\order{\as^2}$ for $n_f=5$.}

\paragraph{Matching fixed order to resummed calculations}

Pure fixed-order expansions are valid from moderate $y$ to large $y$
($\as \ln^2 y \ll 1$), while resummed calculations apply to small $y$
($y \ll 1$). To obtain predictions over the whole kinematical range it
is necessary to \emph{match} the two calculations. This involves
adding the two calculations and subtracting off double counting. This
alone is not sufficient because even after the subtraction of double
counting, there remain terms from the $\order{\as^2}$ contribution, in
particular a piece $G_{21} \asb^2 L$, which would cause the matched
event-shape distributions to have a $G_{21} \asb^2/y$ divergence at
small $y$.  In contrast, the physical requirement is that the
distribution should vanish at least as fast as a positive power of
$y$. The matching procedure is therefore more involved than a
simple subtraction of double-counting terms between the resummed and
fixed-order contributions; the details are given below.

%
%

Matching can be performed either for $R$ or the logarithm of $R$,
the resulting expressions are identical to \oaa, but differ in
the treatment of subleading terms.  The prediction of the {\bf Log(R)}
matching scheme is given by \cite{CTTW}:
\begin{eqnarray}\label{eq:logR}
\ln R(y,\alpha_s) & = & L g_1(\alpha_s L) +g_2(\alpha_s L) - (G_{11}L +G_{12}L^2)\alphab \\
        & - &(G_{22}L^2 +G_{23}L^3)\alphab^2
          + {\cal A}(y)\alphab +\left[{\cal B}(y)- \frac{1}{2}{\cal A}^2(y)\right]\alphab^2 \; .  \nonumber
\end{eqnarray}
As the entire $\order{\as^2}$ term, ${\cal B}(y)$, is exponentiated by
this procedure, the problem of unphysical divergence from
the $G_{21}$ term is avoided.

The expression for the {\bf R} matching scheme reads \cite{CTTW}
\begin{eqnarray}\label{eq:R}
R(y,\alpha_s) & = & (1+C_1 \alphab + C_2\alphab^2)
\exp{\left[L g_1(\alpha_s L) +g_2(\alpha_s L) + G_{21}L\alphab^2\right]}  \\ \nonumber
        & - & G_{21}L\alphab^2 - \left[C_1 + G_{11}L +G_{12}L^2\right]\alphab\\ \nonumber
 & - & \left[C_2 + C_1(G_{11}L +G_{12}L^2) + \frac{1}{2}(G_{11}L +G_{12}L^2)^2 + (G_{22}L^2 +G_{23}L^3)\right]\alphab^2\\ \nonumber
  & + & {\cal A}(y)\alphab +{\cal B}(y)\alphab^2  \; .\nonumber
\end{eqnarray}
Here the $G_{21}$ term is explicitly placed in the exponent, with the
non-exponentiated fixed-order remainder vanishing as $y$ is taken to
zero.

The Log(R) scheme is generally preferred over the R scheme, because
the latter requires explicit knowledge of $G_{21}$ and $C_2$, which
have to be evaluated numerically. The Log(R) expression is furthermore
much simpler and it is believed to be theoretically more stable. This
can be seen in the region of small $y$, where the remainder terms of
the R scheme are found to be larger than the corresponding Log(R)
terms.

\subsection{Sources of arbitrariness}

\paragraph{Modified matching}
The predictions obtained with the Log(R) and R matching schemes
suffer from a limitation: unlike fixed-order predictions, they do
not vanish at the multi-jet kinematic limit. The $\oaa$
expressions vanish at the phase space limit for four-parton
production. The matching schemes can be modified to overcome this
drawback. To do this, a kinematic constraint is imposed to
guarantee that the prediction of the distribution vanishes at a
given value $\ymax$. This means for the {\bf modified Log(R)}
\cite{CTTW}
\begin{equation}\label{eq:modlogR}
\ln R(\ymax) = 0 \; , \;  \left.\frac{1}{\sigma_{tot}}\; \frac{d\sigma(y)}{dy}\right|_{y=\ymax} = \left.\frac{dR}{dy}\right|_{y=\ymax}=0 \; .
\end{equation}
To fulfil this constraint $L$ is replaced by
\begin{equation}
\tilde{L}  = \frac{1}{p}\ln\left( \left(\frac{y_0}{y}\right)^p - \left(\frac{y_0}
{\ymax}\right)^p +1\right) \; .
\end{equation}
The power $p$ is called the degree of modification and is usually
chosen equal to unity.  It determines how fast the distribution is
damped at the kinematic limit.  The nominal values of $\ymax$ are
obtained for thrust, $C$-parameter and $y_3$ on the basis of symmetry
arguments. For the other variables matrix element calculations were
carried out and compared to various parton shower simulations using
PYTHIA \cite{pythia}, the results of which are given in
Table~\ref{tab:ymax}. Ten million events were generated for the Monte Carlo
simulation.  Since the aim behind the modified matching formulae is to
extend the distribution up to the true kinematic maximum of the
observable, the maximum of all $\ymax$ determinations is taken as
nominal $\ymax$ value. However, insofar as the prescription for
this extension is quite arbitrary, for studies of the theoretical
uncertainties it will also be instructive to examine, as an
alternative for each observable, the lowest of the $y_{\max}$ values
found in Table~\ref{tab:ymax} (see Section \ref{sec:error}).
\TABULAR[h]{|l|l|c|c|c|c|c|c|}{ \hline variable  & $\tau$ &
$\rho$ & $-\ln y_3^\ast$ & $  B_T $ & $ B_W $ & $C$ & comment  \\
\hline $\ymax$ & 0.5    &  --   & $\ln 3$ &  --   & --  & 1  & theoretical maximum \\
\hline $\ymax$ & 0.4225 & 0.4175 & $1.100$ &  0.4075 & 0.325 & 1  & ME 4-partons \\
\hline $\ymax$ & 0.428 & 0.394  & $1.102$ &  0.397 & 0.307 & 0.994  & PS partons \\
\hline $\ymax$ & 0.434 & 0.383  & $1.103$ &  0.396 & 0.295 & 0.995  & PS hadrons \\ \hline
\hline $\ymax$ & 0.5    & 0.42   & $\ln 3$ &  0.41   & 0.33  & 1  & nominal value \\
\hline $y^\prime_{\rm max}$ & 0.42    & 0.38   & $\ln 3$ &  0.39   & 0.29
& 0.99  & alternate lower value \\
\hline

} {\label{tab:ymax}Values of $\ymax$ at which the distributions
vanish. The `ME 4-partons' entries have been determined from
binned distributions and their accuracy is therefore limited by
the resolution of the binning. $^\ast$In the case of $-\ln y_3$ it
is the minimal value, corresponding to $y_{3,max}=1/3$.}

While for the modification of Log(R)-matching the replacement of $L$
with $\tilde{L}$ in Eq.~(\ref{eq:logR}) is sufficient to fulfil the constraints
of Eq.~(\ref{eq:modlogR}) this
is not true for the R-matching. To modify the R-matching in addition,
the matching coefficients $G_{11}$ and $G_{12}$ become functions of $y$
such that:
\begin{equation}
\tilde{L}(\ymax) = 0 \; , \; \tilde{G}_{11}(\ymax) = 0  \; , \; \tilde{G}_{21}(\ymax) = 0 \; .
\end{equation}
This is achieved with the following modification:
\begin{eqnarray}\label{eq:modified}
\tilde{L}(y) & = &  \frac{1}{p}\ln\left[ \left(\frac{y_0}{y}\right)^p - \left(\frac{y_0}
{\ymax}\right)^p +1\right]\; , \\ \nonumber
\tilde{G}_{11}(y) & = & G_{11} \left[1 - \left(\frac{y}{\ymax}\right)^p\right]\; , \\ \nonumber
\tilde{G}_{21}(y) & = & G_{21} \left[1 - \left(\frac{y}{\ymax}\right)^p\right] \; .
\end{eqnarray}
%
Finally the expression for the {\bf modified R} matching scheme can be written as
\begin{eqnarray}\label{modR}
\tilde{R}(y,\alpha_s) & = & (1+C_1 \alphab + C_2\alphab^2) \\ \nonumber
& & \times \; \exp{\left[\tilde{L} g_1(\alpha_s \tilde{L}) +g_2(\alpha_s \tilde{L}) -
 \left(\frac{y}{\ymax}\right)^p G_{11}\alphab\tilde{L}+
\tilde{G}_{21}\tilde{L}\alphab^2\right]}  \\ \nonumber
        & - & \tilde{G}_{21}\tilde{L}\alphab^2 - \left[C_1 + \tilde{G}_{11}\tilde{L} +G_{12}\tilde{L}^2\right]\alphab\\ \nonumber
 & - & \left[C_2 + C_1(\tilde{G}_{11}\tilde{L} +G_{12}\tilde{L}^2) + \frac{1}{2}(\tilde{G}_{11}\tilde{L} +G_{12}\tilde{L}^2)^2 + (G_{22}\tilde{L}^2 +G_{23}\tilde{L}^3)\right]\alphab^2\\ \nonumber
  & + & {\cal A}(y)\alphab +{\cal B}(y)\alphab^2  \; .\nonumber
\end{eqnarray}
\paragraph{Renormalisation scale dependence}
For scale parameters $\xmu$ different from unity, every second order
terms acquires a scale dependence explicitly given by
\begin{eqnarray}\label{eq:xmu}
{\cal B}(y) & \rightarrow &  {\cal {\overline B}}(y) = {\cal B}(y) + 2 \beta_0 \, {\cal A}(y)\ln\xmu^2 \; ,\\ \nonumber
G_{21} & \rightarrow &  {\overline G}_{21} = G_{21} + 2 \beta_0 \, G_{11}\ln\xmu^2 \; , \\ \nonumber
G_{22} & \rightarrow &  {\overline G}_{22} = G_{22} + 2 \beta_0 \, G_{12}\ln\xmu^2 \; , \\ \nonumber
C_{2} & \rightarrow &  {\overline C}_{2} = C_{2} + 2 \beta_0 \, C_{1}\ln\xmu^2 \; , \\ \nonumber
g_{2}(\alpha_s L) & \rightarrow &  {\overline g}_{2}(\alpha_s L) = g_{2}(\alpha_s L) +
\frac{\beta_0}{\pi}\left(\alpha_s L \right)^2 \, g_{1}^\prime(\alpha_s L)\ln\xmu^2 \; . \\ \nonumber
\end{eqnarray}
The overlined terms of Eq.~(\ref{eq:xmu}) replace for $\xmu \neq 1$
the corresponding terms in equations (\ref{eq:logR}) and (\ref{eq:R}).
Of course the coupling constant itself exhibits the scale dependence
indicated in Eq.~(\ref{running_formula}).
\paragraph{Rescaling resummed logarithms}

In addition to the arbitrariness in the choice of $\mu$ there is
also arbitrariness in the definition of the logarithms to be
resummed; for example, whether powers of $\as \ln \frac{y_0}{y}$
or of $\as \ln \frac{2 y_0}{y}$ are resummed. This can be
formalised \cite{DISXL} by the introduction of an $x_L$ parameter,
analogous to $x_\mu$, such that where normally powers of $\as \ln
\frac{y_0}{y}$ are resummed instead, powers of $\as \ln
\frac{y_0}{x_L y}$ are resummed. Such a rescaling alters the
resummed formulae in the modified case according to :
\begin{eqnarray}
 \tilde L & \to &
   \widehat L = \frac{1}{p}\ln \left[\left(\frac{y_0}{x_L \cdot y}\right)^p
          - \left(\frac{y_0}{x_L \cdot \ymax}\right)^p +
    1\right] \, , \\
  g_1(\as L) & \to & {\widehat g}_1 = g_1(\as \widehat L)\,, \\
  g_2(\as L) & \to & {\widehat g}_2 = g_2(\as \widehat L) + \ln x_L
  \frac{d}{d\widehat L} \left(
    \widehat L g_1(\as \widehat L) \right)\,,
\end{eqnarray}
where the quantity $-\frac{d}{d\widehat L} \left( \widehat L g_1(\as
  \widehat L) \right)$ is referred to in some contexts as $R'(\as
\widehat L)$.
Rescaling the argument of the logarithm also entails changes to the fixed-order coefficients
both in the modified and unmodified cases:
\begin{eqnarray}
  G_{12} & \to & \widehat G_{12} = G_{12}\\ \nonumber
  G_{11} & \to & \widehat G_{11} = G_{11} + 2 G_{12} \ln x_L \\ \nonumber
  G_{23} & \to & \widehat G_{23} = G_{23} \\ \nonumber
  G_{22} & \to & \widehat G_{22} = G_{22} + 3 G_{23} \ln x_L \\ \nonumber
  G_{21} & \to & \widehat G_{21} = G_{21} + 2G_{22} \ln x_L + 3 G_{23}
\ln^2 x_L \\ \nonumber
  C_1 & \to & \widehat C_1 = C_1 + G_{11} \ln x_L + G_{12} \ln^2 x_L
\\\nonumber
  C_2 & \to & \widehat C_2 = C_2 + (C_1 G_{11} + G_{21})\ln x_L
                          + (C_{1}G_{12}+G_{22}+\half G_{11}^2)\ln^2 x_L\\\nonumber
                        &  & \qquad\qquad\qquad\qquad\nonumber
                          + (G_{23}+G_{12}G_{11})\ln^3 x_L
                          +  \half G_{12}^2\ln^4 x_L \; .
\end{eqnarray}
Transformations of the expressions under $\xmu$ and $x_L$ variations
are commutative and can therefore be carried out in any order. In the case of
the modified R matching scheme, factors of the type $\left[1 -
  \left(\frac{y}{\ymax}\right)^p\right]$ are to be applied to
$\widehat G_{11}$ and $ \widehat G_{21}$ (Eq.~\ref{eq:modified}) after
the $x_L$ variation.

\subsection{Further estimates}

\paragraph{Direct estimates of higher-orders} %
It is to be noted that for the thrust and heavy jet mass,
investigations have been carried out of potential sources of
higher order terms (NNLL, etc.) in the resummation, in particular
those associated with the running of the coupling, using the
dressed-gluon exponentiation model \cite{gardi}. These estimated
higher orders have a rather large effect on fits for $\as$,
somewhat larger than the uncertainties which are deduced in
Section \ref{sec:error}. Given that these are strictly speaking
only model calculations and that they exist for only a subset of
the observables studied here (the thrust, heavy jet mass and
$C$-parameter), they are not included here in the uncertainty
estimates; their existence should however be kept in mind,
together with the possibility that `standard' methods (e.g.\ scale
variations) for estimating the size of higher-order effects may be
overly optimistic.

\paragraph{Heavy-quark effects} %
At $\mz$, events with primary b quarks represent about $20\%$ of all
events. The theoretical calculations assume however light quarks. It
is important therefore to understand the impact of heavy quarks on the
theoretical predictions. Fixed order predictions with heavy quarks
have been in existence for a few years \cite{HQNLO} and distributions
with b quarks are known to differ by a few
percent from light-quark distributions. Once this effect is multiplied
by the fraction of b quark events it becomes of the order of a
percent \cite{ALEPH_masseffect}, which is small relative to the other perturbative uncertainties.

Recently the first NLL resummed calculation for jet rates in
heavy-quark events was completed \cite{Krauss} (though it is NLL for
$R$ as opposed to $\ln R$, {\it i.e.}\ a lesser accuracy than that used
throughout this paper). Physically there are two effects at play.
Firstly there is the `dead cone' (the suppression of collinear emissions with
an angle smaller than $m_b/Q$) which implies a modification of the
double-logarithmic structure for $y_3$ below
a critical value, $y_{3c}$, of order $m_b^2/Q^2$:
\begin{equation}
\label{eq:HeavyQuarkDead}
G_{12} \ln^2 \frac1{y_3} \to G_{12} \left( \ln^2 \frac1{y_3}
 - \ln^2 \frac{y_{3c}}{ y_3} \right)\,\qquad\quad (y_3 \lesssim y_{3c})\,.
\end{equation}
Secondly higher order terms are modified because the number of
active flavours (e.g. in $\beta_0$) decreases by one for the parts
of the momentum integral with transverse momenta less than $m_b$
(corresponding also to $y_3 < y_{3c}$). The authors of
\cite{Krauss} quote effects for jet rates of the order of a 3--4\%
for $y_{\rm cut}=0.004$ with $m_b=5$~GeV and $Q=\mz$.

Currently no calculations exist for other observables. However the same
physical arguments allow one to make the statement that the dead cone will
lead to a modification of the double logarithms analogous to that of
\eqref{eq:HeavyQuarkDead}, below a critical value $y_c \sim m_b^n/Q^n$, with
$n=1$ for the broadenings and $n=2$ for the thrust and $C$-parameter
(the heavy jet mass is more complicated because of the direct b-quark mass
contribution). In addition, for the thrust, jet-mass and $C$-parameter, in
the range $m_b/Q \lesssim y \lesssim m_b^2/Q^2$, there is a mixture in the
resummation of contributions with $4$ and $5$ active flavours.

\paragraph{NNL calculations} %

One source of future improvement in the theoretical accuracy is
expected to come from calculations of higher order contributions.
Considerable progress has been achieved in calculating two-loop
amplitudes for the process $\epem \rightarrow {\rm q \bar q g}$
\cite{2-loop}. Subtraction methods at NNLO to cancel infrared and
collinear divergences between the two-loop, one-loop and tree-level
contributions are currently being developed (see for example
\cite{Weinzierl}). It will then be necessary to combine the various
elements in the form of a fixed-order Monte Carlo program (analogous
to EVENT2 \cite{EVENT2}), which can be used to calculate the
$\order{\as^3}$ contribution to the event-shape distributions. This is
expected to reduce the scale dependence (see for example
\cite{Glover}), especially in the three-jet region. In the two-jet region,
the gain from the NNLO calculations could be more modest, because much
of the $\as^3$ contribution is already embodied in the NLL
resummation. Improved accuracy over the full phase-space may therefore
also require a NNLL resummed calculation. Progress on such
calculations is also being made, though full results have so far been
obtained only for observables that are somewhat simpler than event
shapes, such as the Higgs $p_t$ distribution at hadron colliders
\cite{BCFG}.


\section{Estimating theoretical uncertainties}\label{sec:error}

For the purpose of $\alpha_s$ measurements it is necessary to adopt a
nominal theoretical prediction, which is to be used to determine the
central value of $\alpha_s$, as well as a set of variations for
estimating the uncertainty on $\as$.

For the matching scheme the use of modified Log(R)
matching is advocated, since it has proven more stable than the modified R matching
scheme, which will be used as the `alternative' theory for uncertainty
estimates.

The default value of $\ymax$ is the maximal possible value for any
number of partons (as obtained from theoretical arguments and
parton-shower simulations, Table~\ref{tab:ymax}), while the lower
limit $y^\prime_{\rm max}$ is used as an extreme alternative estimate.

The options for the modification degree $p$ are less clearly
delimited --- properties of certain fixed-order calculations
\cite{tmresum,CTTW,newbresum} suggest that $p$ should be $\ge 1$.
The simplest case $p=1$ is recommended for the nominal
configuration. The effect of the value of $p$ on the extracted
value of $\alpha_s$ is studied by fitting predictions with
$\alpha_s$ as free parameter and $p\neq 1$ to the reference theory
with $p=1$, using weights proportional to the differential cross
section. The change in $\alpha_s$ is depicted in
Fig.~\ref{fig:pvary}. \EPSFIGURE[ht!]{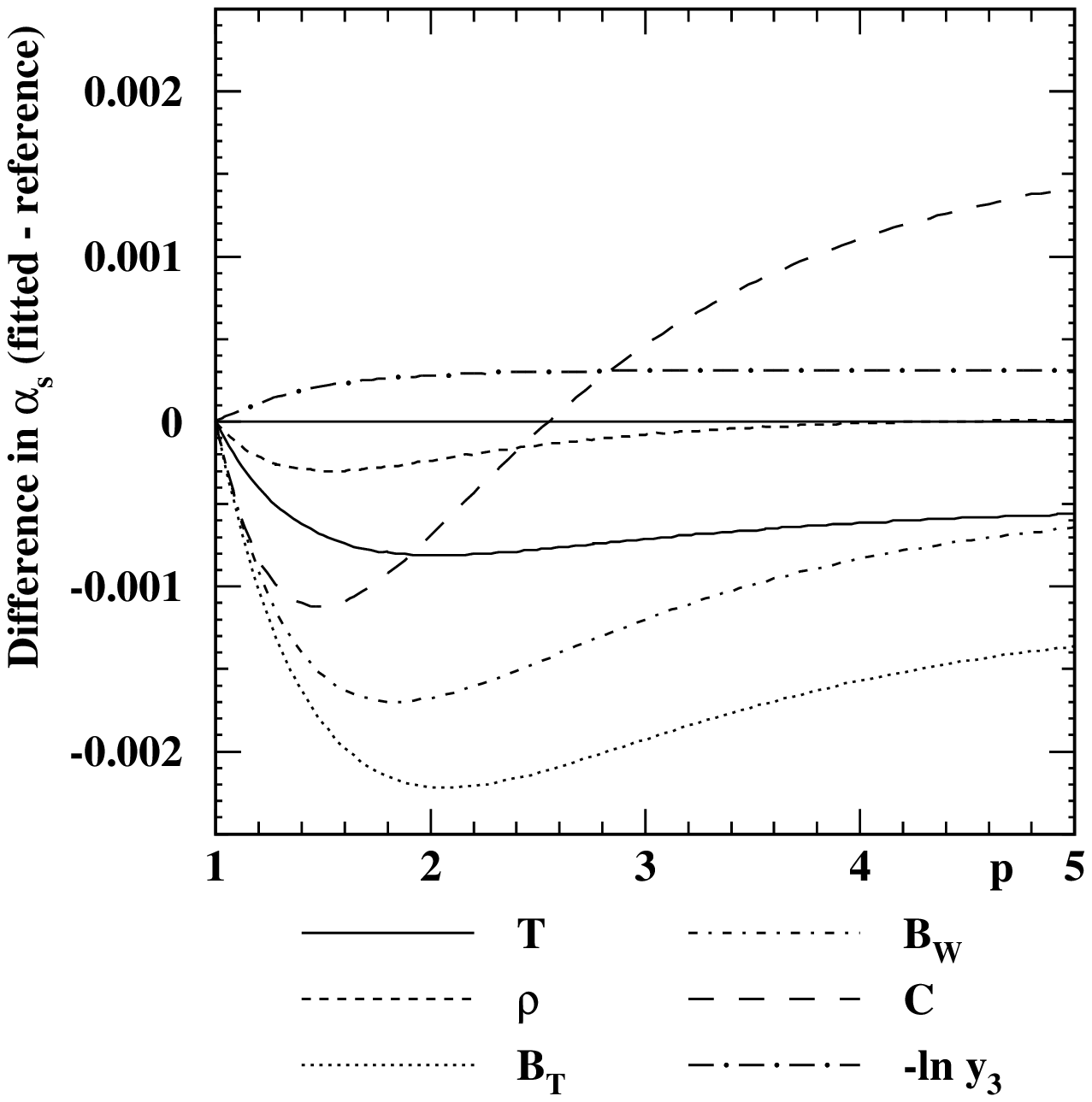,width=130mm}
{The change in extracted $\alpha_s$ as a function of the degree of
modification $p$ for several observables. As reference theory the
modified Log(R) matching scheme is taken with $p=1$. Predictions
with different $p$ are fit to the reference with $\alpha_s$ as
free parameter.\label{fig:pvary}} In the limit of large $p$ the
prediction turns into the disfavoured unmodified matching scheme.
In general the largest difference in $\alpha_s$ is observed around
$p=2$, which is suggested for systematic purposes.

For the renormalisation scale, in a fixed order framework $\mu^2 =
Q^2$ is the simplest, hence natural choice for the invariant scale of
the process.  So in order to enable straightforward comparison with
existing measurements the conventional $\xmu=1$ is recommended,
together with the standard variation range, $1/2<\xmu <2$. The
renormalisation scale is kept the same in the fixed order and
resummation parts of the calculation --- though not strictly required
this is also conventional, and varying $\xmu$ separately in the two
parts of the calculation would complicate the formalism somewhat.

A novel type of systematic study is proposed for the proper
resummation part of the theoretical prediction. The arbitrariness
of the logarithmic terms to be resummed to all orders is
formalised through the variable rescaling factor $x_L$. In analogy
with the situation for renormalisation scale dependence, if a
resummation to all orders of logarithmic accuracy (LL, NLL, NNLL,
\ldots) was complete then the predictions would be independent of
the choice of $x_L$. But at any truncated order (e.g.\ NLL) there
is a residual $x_L$ dependence, which like the $x_\mu$ dependence,
can be used to gauge the expected order of magnitude of missing
higher order contributions.  As in the case of the $\xmu$ scale
dependence, the range of scales can not be derived from first
principles. One of the most critical elements in the measurement
of $\alpha_s$ and estimation of the associated uncertainties is
the choice of a default value and range of variation for $x_L$.
All existing $\epem$ calculations implicitly use $x_L=1$ and this
convention has always been assumed for measurements of $\alpha_s$.
It is possible to argue for this as the most sensible default
choice on the following grounds. In the resummation procedure, one
evaluates integrals over the rapidities $\eta$ and transverse
momenta $k_T$ of gluons. With the values of $y_0$ given after
Eq.~(\ref{eq:Resummed35}), one can show that the value of the
observable in the presence of a single soft and collinear gluon
emission is $y\simeq y_0(k_T/Q)^a \cdot \exp(-b \eta)$, where $a$
and $b$ are integers. Accordingly the logarithm that is resummed,
$L\equiv \ln y_0/(x_L y)$, can be rewritten $a \ln Q/k_t + b\eta -
\ln x_L$. It then appears quite natural to choose the convention
$x_L=1$, since $L$ reduces to the combination of the physical
logarithms, $a \ln Q/k_t + b\eta$, without any extra constant
piece.

While there exists a simple motivation for the choice of a central
value of $x_L=1$, implicitly embodied in current standard practice,
the choice of the range of $x_L$ is far more subjective and has never
been considered so far. An ad hoc prescription would be to vary
$x_L$ in the same canonical range as $\xmu$. The effect of a given
value of $x_L$ on the distribution must be studied thoroughly,
however, the objective being to find an estimate of missing higher
orders. A reasonable range for $x_L$ variations should not
over-estimate the theoretical uncertainty, but complement the
other investigations. Even if a strict range setting is
impossible, a sensible proposal will be elaborated on the basis of
various tests.
This issue is studied in detail in the next
subsection.

Once all these different sources of uncertainty have been examined,
it is necessary to combine them, bearing in mind that there is only
partial complementarity between them. This is achieved with the uncertainty band method,
which is discussed in Section~\ref{sec:UncertaintyBand}.

\subsection[Setting a range for $x_L$]{Setting a range for $\boldsymbol{x_L}$}
The determination of range of variation for $x_L$ is quite
critical, because its effect is rather large. One approach is to
try and find some theoretical motivation for a range, which leads
to a number of possibilities:
\begin{itemize}
\item The observation that many of the theoretical
  calculation involve an inverse Mellin transform that leads naturally
  to logarithms of $e^{\gae}/y$, suggesting a range $| \ln x_L| <
  \gae$, where $\gae \simeq .57721566$. This leads to $0.56 \leq x_L \leq 1.78$.

\item Certain values of $x_L$ lead some of the subleading
fixed-order
  expansion coefficients being zero; for example $\ln x_L = \frac34$
  (twice this for $-\ln y_3$) gives $\Govrln_{11} = 0$,
  suggesting $| \ln x_L| < \frac34$, {\it i.e.} $0.47 \leq x_L \leq 2.12$
 (double for $-\ln y_3$). The appearance
  of a doubled range for $-\log y_3$ is natural also with the observation that
  for a single emission, the two-jet resolution parameter is the
  square of the jet broadening.

\item If instead  $\Govrln_{21}$ is wanted to be zero then a
  quadratic equation has to be solved for $\ln x_L$, giving the results shown in
  Table~\ref{tab:xrange}. Taking the solution closer to zero as more
  natural suggests an average range of about $|\ln x_L| \lesssim 0.6$,
{\it i.e.}
  $0.54 \leq x_L \leq 1.86$.
\end{itemize}
\begin{table}[htbp]
  \begin{center}
    \begin{tabular}{|l|c|c|}\hline
      Observable & $(\ln x_L)_-$ & $(\ln x_L)_+$ \\ \hline
      $T$ & $-1.98$ & $0.36$ \\ \hline
      $C$ & $-2.39$ & $0.76$ \\ \hline
      $B_T$ & $-2.00$ & $0.48$ \\ \hline
      $B_W$ & $-1.15$ & $0.78$ \\ \hline
      $-\ln y_3$ & --- & --- \\ \hline
      $\rho$ & $-1.60$ & $0.73$ \\ \hline
    \end{tabular}
  \end{center}
  \caption{Values of $\ln x_L$ which give $\Govrln_{21}=0$. The results
    typically have an error of about $\pm 0.05$ because $G_{21}$ is
    known only from fits to fixed-order Monte Carlo results. In the
  case of $-\ln y_3$ the solutions are complex.}
  \label{tab:xrange}
\end{table}

Another approach consists of a comparison of the $\oaa$
calculation with the expansion to second order of the resummed
NLL prediction. The latter is obtained by expanding Eq.~(\ref{eq:RNLL}),
keeping only terms up to $\oaa$.  The difference
between these two expression is sensitive to asymptotic terms
present in the exact $\oaa$ calculation but absent in the NLL
expansion. This Ansatz conserves the information of the
differential distribution, it can be expected that in general the
uncertainties are not constant across the spectra.
The difference $\Delta(y)$ between the
NLL expansion and the exact $\oaa$ calculation is determined for
a central value of $\alpha_s=0.12$. Then a theoretical variation
is constructed by adding (resp. subtracting) $\Delta(y)$ to the
reference prediction ({\it i.e.} using the modified Log(R) matching scheme with
$x_L = 1$). Finally the reference theory is used with variable
$x_L$ as free parameter to fit the variation. In practice, three
different fit ranges, given in Table~\ref{tab:xlfitrange},
are used to test the stability of the procedure. The first range (nominal)
covers experimental fit ranges for $\alpha_s$, the second range (2-jet) is
restricted to the semi-inclusive region and the third range (3-jet) comprises
multi-jet production.
\TABULAR[h]{|l|c|c|c|}
{
\hline
Observable & fit range 1 (nominal) & fit range 2 (2-jet) & fit range 3 (3-jet)  \\ \hline
$T$        & $0.70-0.97$ & $0.87-0.97$ & $0.70-0.83$ \\ \hline
$-\ln y_3$ & $2.6-7.2$   & $5.0-7.2$   & $2.6-5.0$   \\ \hline
$\rho$     & $0.03-0.25$ & $0.03-0.12$ & $0.14-0.25$ \\ \hline
$B_W$      & $0.04-0.20$ & $0.04-0.12$ & $0.12-0.20$ \\ \hline
$C$        & $0.08-0.70$ & $0.08-0.30$ & $0.38-0.70$ \\ \hline
$B_T$      & $0.05-0.28$ & $0.05-0.14$ & $0.15-0.28$ \\ \hline
}
{\label{tab:xlfitrange}Fit ranges used for the determination
of upper and lower bounds of $x_L$. }

For the fitting procedure statistical weights scaling with the
square root of the distribution value are applied. The resulting
limits on $x_L$ are given in Table~\ref{tab:xlexp}.  It turns out
that they strongly depend upon the range of the fit: a variation
of $x_L$ leads to a change in shape which is very large below the
peak region in the two-jet limit. The effect is minimal around the
peak and then increases continuously towards $\ymax$.  The shape
of $\Delta(y)$ is similar, but its slope is much steeper.  This is
reflected in the dependence of the results on the fit range. The
resummation technique and the evaluation of $\Delta(y)$ are
generally considered to be applicable in the semi-inclusive
region, substantiating results obtained with the two-jet fit
range.  As a cross-check the same procedure is applied to the case
of $\xmu$. In average and for the two-jet fit range a span for
$\xmu$ from 0.4 to 2.9 is found, in reasonable agreement, although
slightly over-estimating the canonical range from 0.5 to 2.
\TABULAR[h]{|l|c|c|c|}
{
\hline
Observable & $x_L$ nominal range & $x_L$ 2-jet range & $x_L$ 3-jet range \\ \hline
$T$        & $0.55-1.89$ & $0.61-1.67$ & $0.49-8.1$ \\ \hline
$-\ln y_3$ & $0.28-2.6$  & $0.29-1.71$ & $0.25-2.86$   \\ \hline
$\rho$     & $0.48-2.17$ & $0.59-1.69$ & $0.27-4.9$ \\ \hline
$B_W$      & $0.36-3.10$ & $0.38-2.70$ & $0.33-5.0$ \\ \hline
$C$        & $0.36-4.40$ & $0.38-3.78$ & $0.35-14.2$ \\ \hline
$B_T$      & $0.50-2.47$ & $0.57-1.99$ & $0.47-7.1$ \\ \hline
}
{\label{tab:xlexp}Ranges of $x_L$ obtained with fits to the difference between
the fixed order $\oaa$ calculation and the second order expansion of
the NLL prediction, for three different fit ranges.}


The comparison with the $\xmu$ variation can be investigated by
determining $x_L$ values such that on average, for a given set of
observables (and fit ranges), the impact of the $x_L$ variation on
a fit of $\as$ is the same as that of the conventional $x_\mu$
variation. While at first sight this may seem to make the $x_L$
variation procedure redundant, it should really be considered as a
method for setting a conventional variation range. Different
observables may then have sensitivities to the common $x_L$ range
that may be (and in practice often are) quite different.
Furthermore, the $y$-shape of the $x_L$ and $x_\mu$ variation will
be different even if their average effect on a fit of $\as$ is the
same, because they probe different subsets of possible
higher-order corrections.


The procedure of this method is as follows: two predictions
are calculated with the modified Log(R) scheme, $\alpha_s=0.12$
and $x_L=1$, one with $\xmu=0.5$, the other with $\xmu=2$. Each of
them are then fitted with the same theory but $\xmu=1$ and $x_L$ being
the free parameter. The results are given in Table~\ref{tab:xlxmu}.
\TABULAR[h]{|l|c|c|c|}
{
\hline
Observable & $x_L$ nominal range  & $x_L$ 2-jet range & $x_L$ 3-jet range \\ \hline
$T$        & $0.56-1.77$ & $0.55-1.78$ & $0.71-1.55$ \\ \hline
$-\ln y_3$ & $0.48-2.23$ & $0.27-2.64$ & $0.59-2.17$  \\ \hline
$\rho$     & $0.58-1.82$ & $0.56-1.86$ & $0.82-1.38$ \\ \hline
$B_W$      & $0.67-1.51$ & $0.63-1.56$ & $0.79-1.35$ \\ \hline
$C$        & $0.61-1.71$ & $0.57-1.81$ & $0.65-1.54$ \\ \hline
$B_T$      & $0.67-1.53$ & $0.65-1.54$ & $0.71-1.46$ \\ \hline
}
{\label{tab:xlxmu}Bounds of $x_L$ obtained with a fit to the
theoretical prediction with $\xmu=0.5$ resp. $\xmu=2.0$.}

Having considered this variety of criteria for choosing the $x_L$
range, it is proposed to set a convention for the $x_L$ range of
$\frac23 < x_L < \frac32$ (equivalently $|\ln x_L| \lesssim
0.405$), which gives an average uncertainty on $\as$ which is
similar in magnitude to that from $x_\mu$ variation.  This is a
slightly narrower range than comes out from the purely theoretical
arguments and from some of the other tests. For the purpose of a
more conservative estimate of the uncertainty, or if one wishes to
consider the uncertainty on the uncertainty, it is suggested to
examine also a wider range $|\ln x_L| < 0.6$. Both ranges will be
used for the numerical estimates in the next section.

\subsection{Uncertainty band method}
\label{sec:UncertaintyBand}
The new method to assess the theoretical systematic uncertainty
for the measurement of $\alpha_s$, called hereafter uncertainty
band method, is composed of two main building blocks:
\begin{enumerate}
\item A nominal reference theory, the modified Log(R) matching scheme,
  used experimentally to determine the value of $\as$.
\item A collection of theoretical uncertainties (variations of the
theory) of the event-shape
  distributions, used to derive the perturbative uncertainty of
  $\as$.
\end{enumerate}
The following variations of the theoretical predictions for the
distributions are taken into account:
\begin{itemize}
\item the renormalisation scale $x_\mu$ is varied  between 0.5 and
2.0,
\item the logarithmic rescaling factor $x_L$  varied in between 2/3 and 3/2 \\
      (for $-\log(y_3)$ an equivalent effect is obtained with squared
      endpoints,
       {\it i.e}. a variation from 4/9 to 9/4),
\item the modified Log(R) matching scheme is replaced by the
modified R matching scheme,
 \item the nominal value of the kinematic
constraint $\ymax$ is
replaced by the lower limit $y^\prime_{\rm max}$ and
\item the first degree modification of the
modified Log(R) matching scheme ($p=1$) is replaced by a second
degree modification ($p=2$).
\end{itemize}

The uncertainty band method gives a direct relation between the
uncertainty of $\alpha_s$ and the uncertainty of the theoretical
prediction. Two pieces of information are required to calculate
the systematic uncertainty: the measured value of the coupling
constant, $\as^0$, and the fit range used for its extraction. With
these elements in hand, the uncertainty can be computed without
re-fitting the data.

The method proceeds in three steps. First the reference
perturbative prediction is calculated for the distribution
in question using a given value for the strong coupling constant $\alpha_s^0$.
Then all variants of the theory
mentioned above are calculated with the same value of
$\alpha_s^0$. In each bin of the distribution, the largest upward
and downward differences with respect to the reference theory are
collected. Hence the uncertainty is set by the extreme values of
the theoretical variants. Variants which lead to similar but
smaller effects are not double-counted. The largest differences
define an uncertainty band around the reference theory.

In the last step, the reference theory is used again, but with
variable $\alpha_s$. In the
spirit of this method, all valid theoretical predictions must lie
within the uncertainty band for the fit
range under consideration. Starting from the nominal $\as^0$, scans of
$\as$ are performed and the validity of resulting predictions is
checked in each bin of the distribution.  The largest and smallest
allowed values of $\alpha_s$ fulfilling the condition are used to finally
set the perturbative systematic error. The method is illustrated for
all variables in Figs.~\ref{fig:leptr}--\ref{fig:lepbt}, with an
input value of $\as^0=0.12$.
It might be argued that the condition for valid
predictions lying strictly inside the uncertainty band is too tight, because
the uncertainty
is basically set by one single bin, which value is subject to statistical
fluctuation of the
numerically computed coefficient functions. An alternative operating mode of the uncertainty
band method consists of a fit of the reference prediction to the uncertainty band
envelope with $\alpha_s$ as free parameter. In this case weights have to be assigned to the
bins inside the fit range. A convenient choice are statistical weights scaling with the square root of the
distribution.

\EPSFIGURE[ht!]{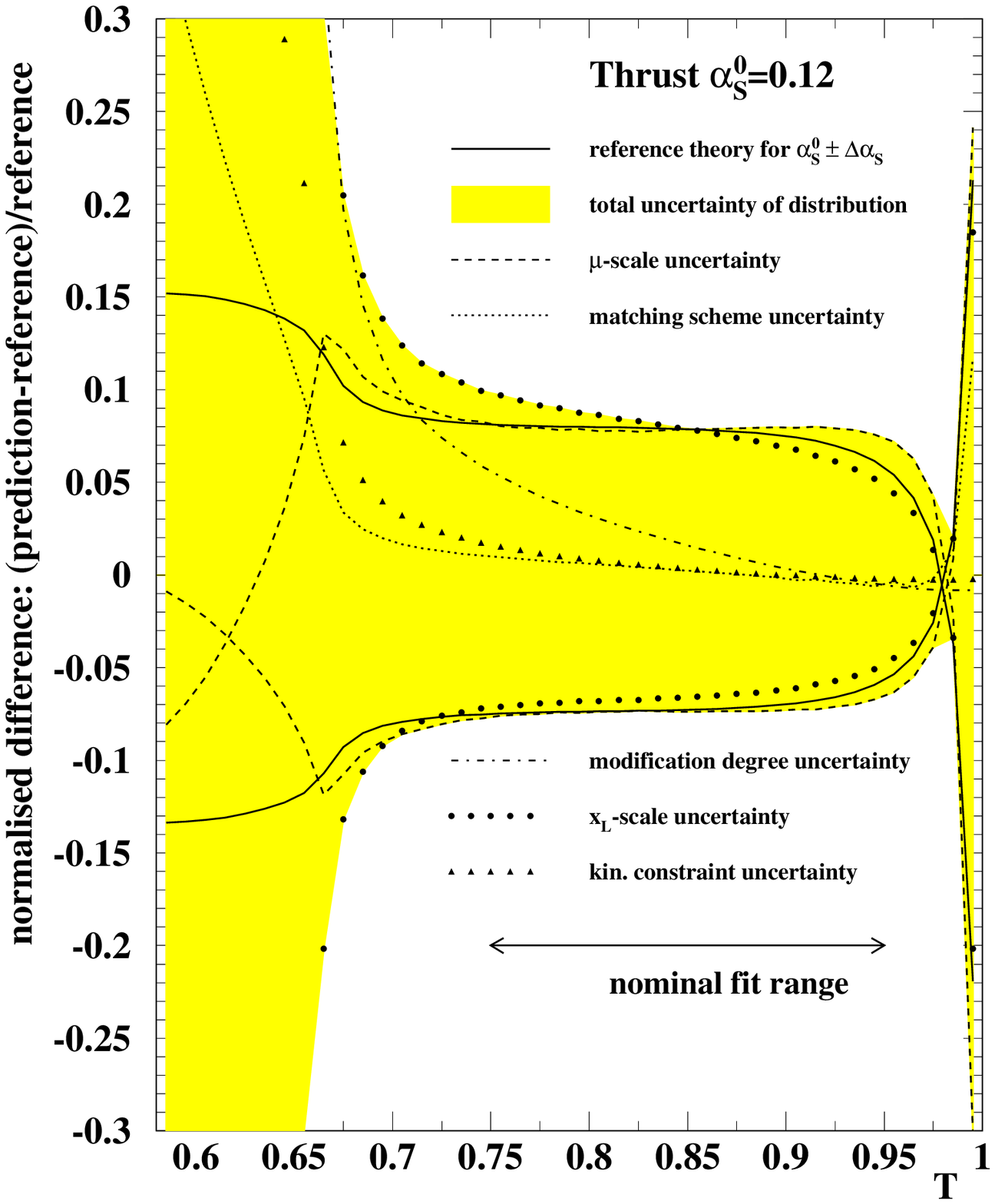,width=130mm}
{Theoretical uncertainties for thrust. The grey area represents the
  total theoretical uncertainty for the distribution. The first curve
  indicates the reference theory with maximum resp.  minimum
  $\alpha_s$, which determines the theoretical uncertainty
  $\Delta\alpha_s$ for a given measurement $\alpha_s^0$. The other
  lines illustrate the different contributions to the theoretical
  uncertainty.\label{fig:leptr}}
\EPSFIGURE[ht!]{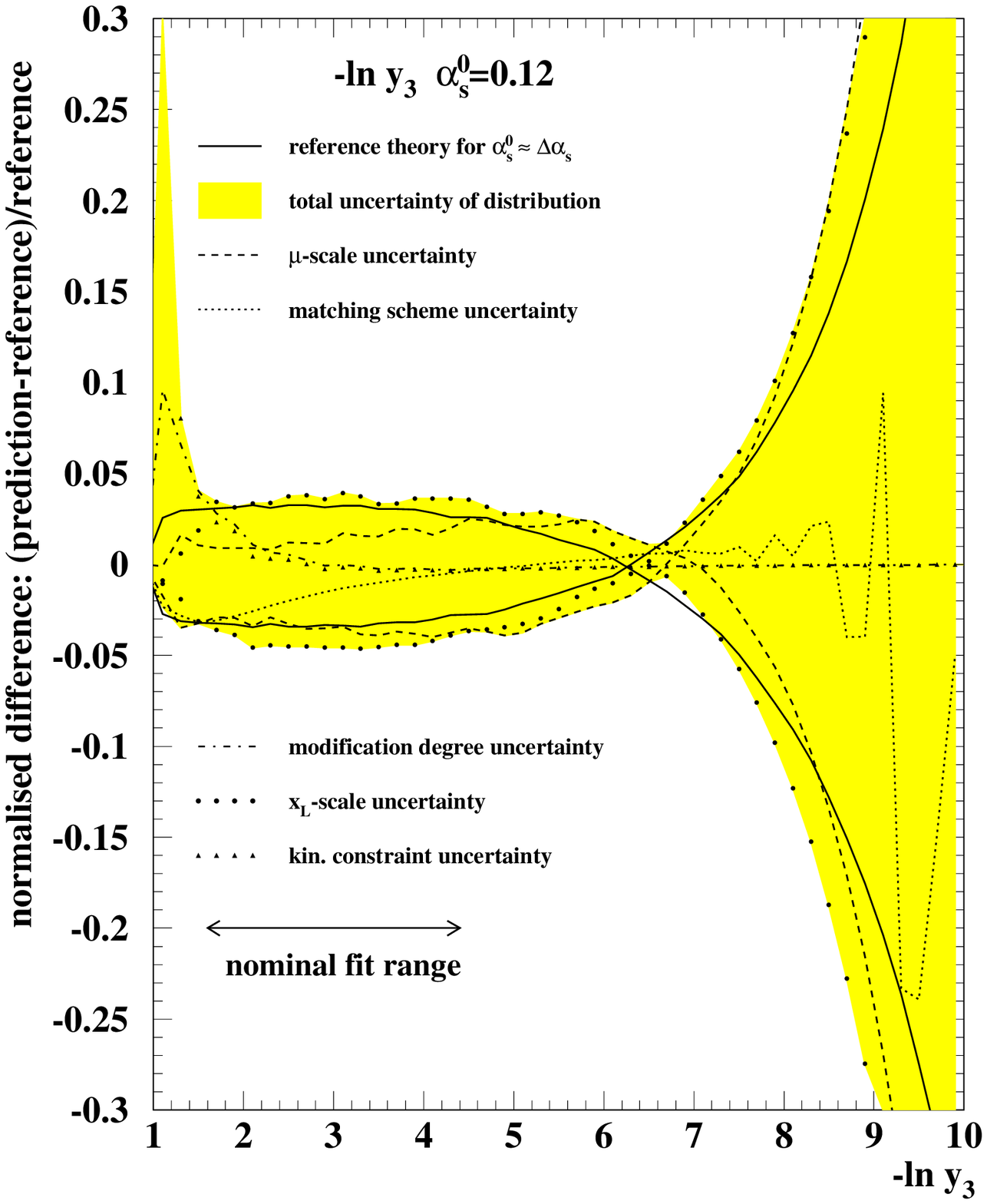,width=130mm} {Theoretical uncertainties for
  $-\ln y_3$. The explanation is given in the caption of
  Fig.~\ref{fig:leptr}. \label{fig:lepy3}}
\EPSFIGURE[ht!]{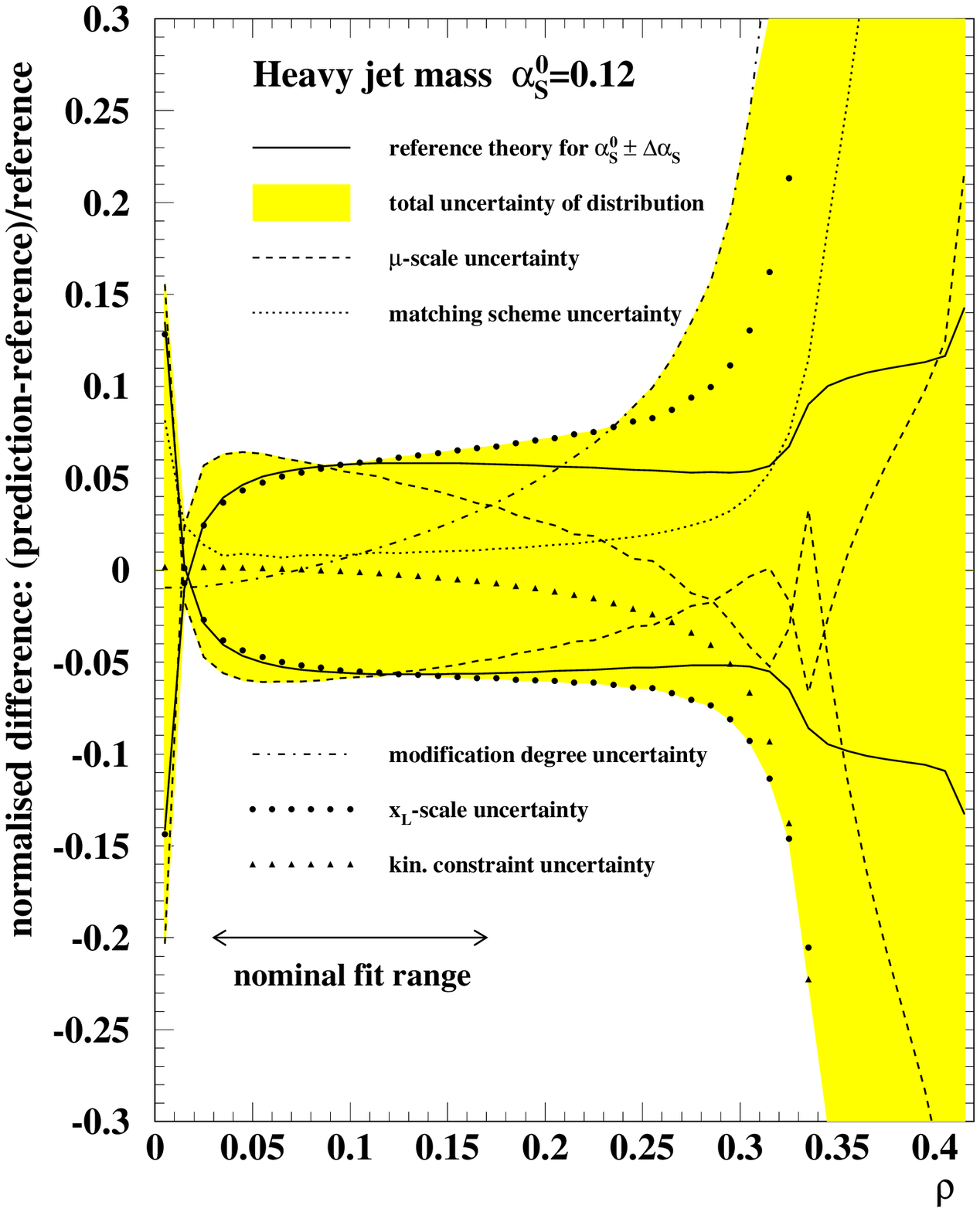,width=130mm} {Theoretical uncertainties for
  heavy jet mass. The explanation is given in the caption of
  Fig.~\ref{fig:leptr}.\label{fig:lepmh}}
\EPSFIGURE[ht!]{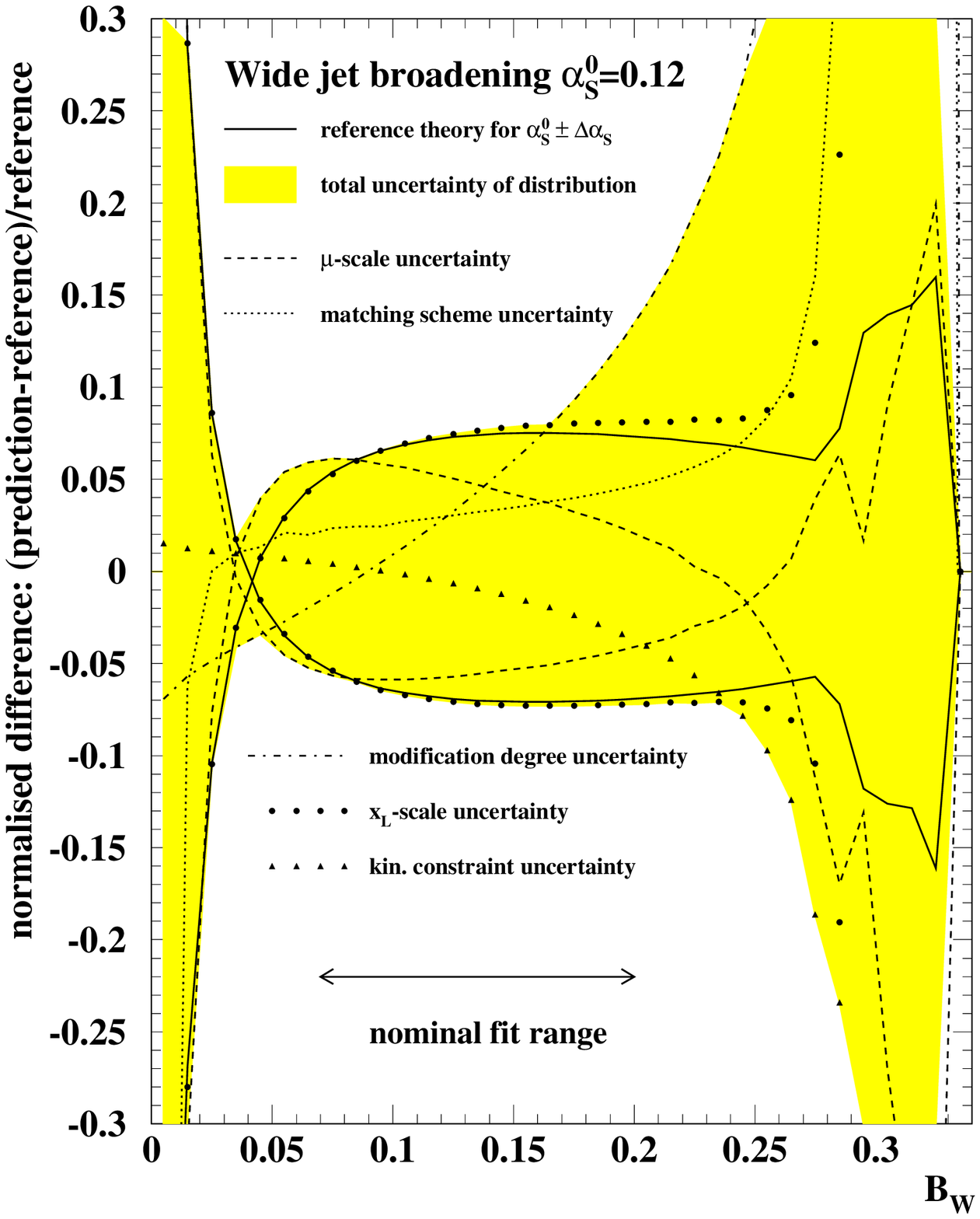,width=130mm} {Theoretical uncertainties for
  wide jet broadening. The explanation is given in the caption of
  Fig.~\ref{fig:leptr}.\label{fig:lepbw}}
\EPSFIGURE[ht!]{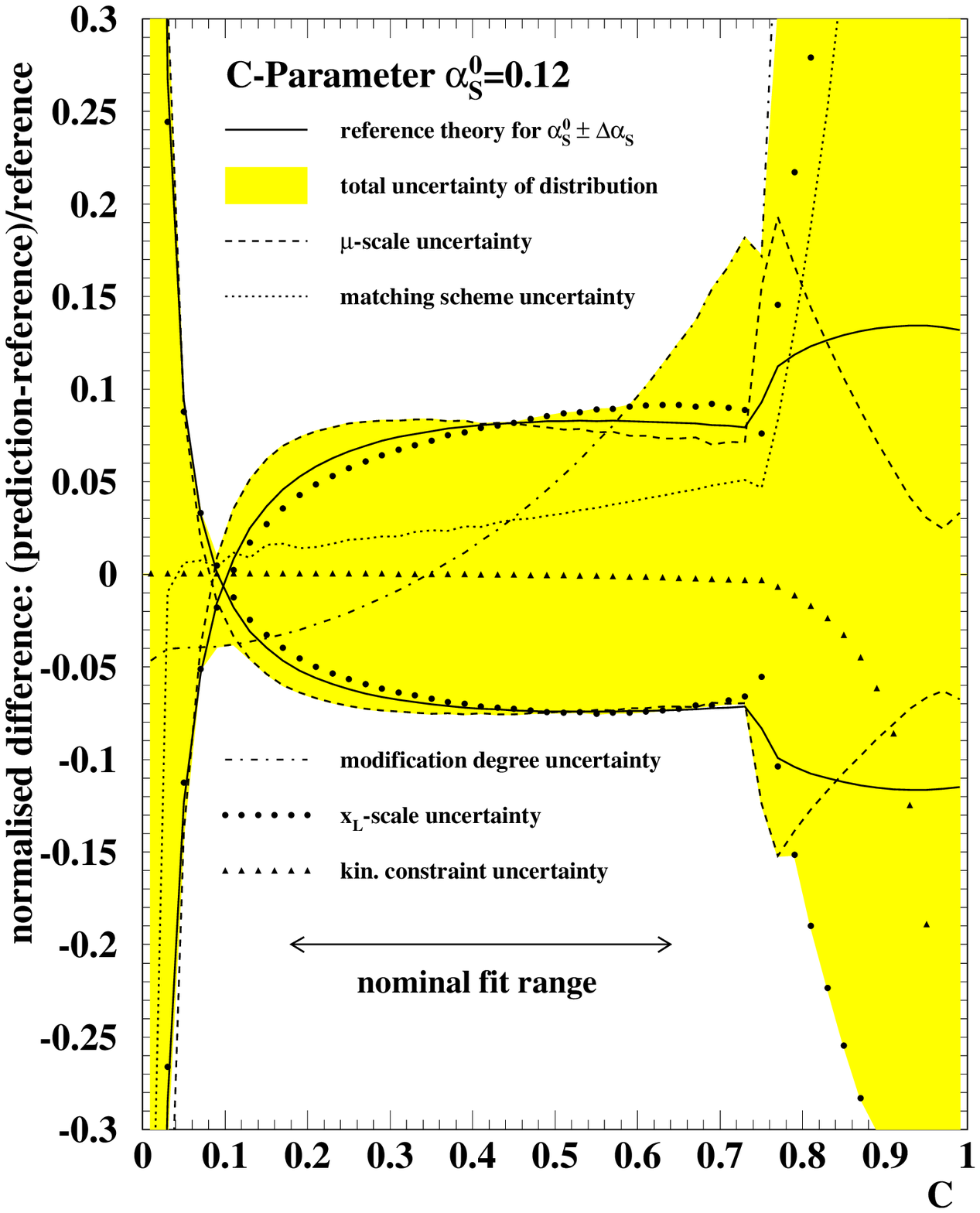,width=130mm} {Theoretical uncertainties for
  $C$ parameter. The explanation is given in the caption of
  Fig.~\ref{fig:leptr}.\label{fig:lepcp}}
\EPSFIGURE[ht!]{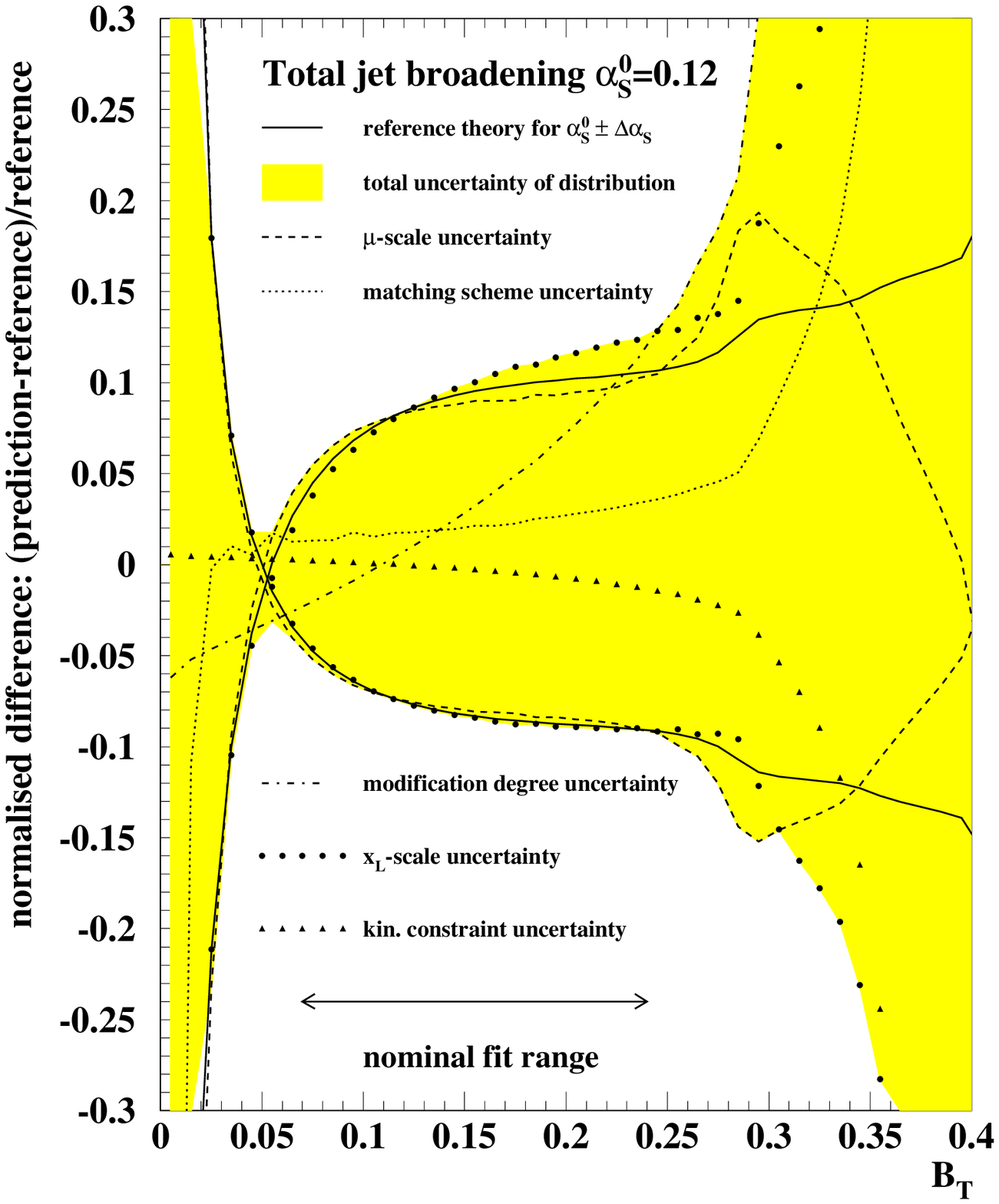,width=130mm} {Theoretical uncertainties for
  total jet broadening. The explanation is given in the caption of
  Fig.~\ref{fig:leptr}.\label{fig:lepbt}}

As can be seen in Figs.~\ref{fig:leptr}--\ref{fig:lepbt}, the total
theoretical uncertainty, {\it i.e.} the uncertainty band, varies drastically
across the distributions. The uncertainties are large, of the order of
30$\%$, at both ends of the spectra and reach a minimum of a few
percent in the peak region. In the range of experimental fits for
$\alpha_s$ the uncertainty of the distribution is between 5$\%$ and 10$\%$.
This turns into an uncertainty for $\alpha_s$ between 3$\%$ and
6$\%$. There is also a considerable spread among the observables,
the uncertainties for the total jet broadening are in the central part
twice as large as for $-\ln y_3$.

A closer look at the individual components of the theoretical
error reveals two major contributions: the variation of the
renormalisation scale and of the $x_L$-scale. In the
semi-inclusive region the $\xmu$ variation generates the dominant
uncertainty while in the three-jet region it is the variation of
$x_L$. These two effects are clearly complementary. Going down in
size of the uncertainty, the degree of modification as probed by
the difference between $p=1$ and $p=2$ is important in the hard
three-jet region, where this effect dominates. Both the kinematic
constraint and the matching scheme uncertainty are rather small in
the central part of the spectra and become somewhat larger at the
multi-jet end.

The uncertainty of $\alpha_s$ is derived from the range of values leading to `valid' predictions
inside the uncertainty band. The resulting uncertainty depends on the fit range, which usually
encloses the central part of the spectrum. The extreme, but still valid reference predictions,
shown as full lines in Figs.~\ref{fig:leptr}--\ref{fig:lepbt}, are in general determined in
the central region, they touch the envelope of the uncertainty band at one point and remain
inside the uncertainty band even outside the fit range.
\clearpage

\section{Results}\label{sec:results}
With the uncertainty band method as tool, all aspects of the
theoretical systematic uncertainties for $\alpha_s$ can be studied in
detail independent of a measured distribution. In Table~\ref{tab:numres} results for the total uncertainty as
well as for individual components are given for a representative fit
range and $\alpha_s=0.12$ as input.
\TABULAR[ht!]{|l|c|c|c|c|c|c|} { \hline & $T$ & $-\ln y_3$ & $\rho$ &
  $B_W$ & $C$ & $B_T$ \\ \hline fit range & 0.78-0.95 & 1.8-4.2 &
  0.05-0.17 & 0.06-0.20 & 0.18-0.62 & 0.07-0.22 \\ \hline
  total       & $+0.0057$ & $+0.0028$ & $+0.0044$ & $+0.0055$ & $+0.0058$ & $+0.0068$ \\
  uncertainty & $-0.0055$ & $-0.0030$ & $-0.0045$ & $-0.0054$ & $-0.0054$ & $-0.0062$ \\ \hline
  $\xmu$      & $+0.0055$ & $+0.0027$ & $+0.0028$ & $+0.0018$ & $+0.0052$ & $+0.0062$ \\
  uncertainty & $-0.0055$ & $-0.0008$ & $-0.0039$ & $-0.0033$ & $-0.0053$ & $-0.0059$ \\ \hline
  $x_L$       & $+0.0048$ & $+0.0017$ & $+0.0041$ & $+0.0053$ & $+0.0044$ & $+0.0045$ \\
  uncertainty & $-0.0047$ & $-0.0029$ & $-0.0042$ & $-0.0053$ & $-0.0046$ & $-0.0059$ \\ \hline
  mod. degree & $-0.0001$ & $+0.0002$ & $-0.0001$ & $-0.0004$ & $-0.0003$ & $+0.0003$ \\
  uncertainty & & & & & & \\ \hline
  matching scheme & $+0.0004$ & $-0.0007$ & $+0.0005$ & $+0.0018$ & $+0.0014$ & $+0.0014$ \\
  uncertainty & & & & & & \\ \hline
  kinematic constraint & $0.0001$ & $-0.0001$ & $-0.0001$ & $-0.0001$ & $+0.0001$ & $+0.0001$ \\
  uncertainty & & & & & & \\ \hline \hline
  total uncertainty     & $+0.0062$ & $+0.0028$ & $+0.0052$ & $+0.0066$ & $+0.0065$ & $+0.0077$ \\
  with $|\ln x_L|<0.6$ & $-0.0059$ & $-0.0032$ & $-0.0052$ & $-0.0066$ & $-0.0061$ & $-0.0073$ \\ \hline
  total uncertainty     & $+0.0062$ & $+0.0032$ & $+0.0052$ & $+0.0061$ & $+0.0066$ & $+0.0074$ \\
  with `fit' method     & $-0.0058$ & $-0.0041$ & $-0.0052$ & $-0.0057$ & $-0.0059$ & $-0.0064$ \\ \hline
} {\label{tab:numres}Theoretical
  uncertainties for $\alpha_s=0.12$ and fit ranges given in the first
  row. For each variable, the total theoretical uncertainty as well as
  uncertainties stemming from individual sources are given. In the last two rows
concern results with more conservative theory assumptions.}
The total uncertainty range from 3$\%$ for $-\ln y_3$ to 6$\%$ for
the total jet broadening, with an overall average of 5$\%$. The
uncertainties related to $\xmu$ and $x_L$ are two-sided and in
general asymmetric, the other uncertainties go in a single
direction. The fact that the total uncertainty is larger than any
of the individual ones is a consequence of the complementary
collection of uncertainties of the distributions in the
uncertainty band. A clear ranking in size of the different
components of the theoretical uncertainty appears for all
variables: variations of $\xmu$ and $x_L$ are most important,
followed by the matching scheme uncertainty, while the kinematic
constraint and modification degree uncertainties are small and of
similar size. Also given in Table~\ref{tab:numres} are uncertainty
estimates obtained with more conservative assumptions, namely a
larger range for $x_L$ ($|\ln x_L|<0.6$) and a fit of the
reference prediction to the uncertainty band envelope (`fit'
method).

Since the uncertainty is calculated with a fixed value of
$\alpha_s$, the dependence of the result on the input value must
be investigated. The input value, normally taken from an
experimental measurement, depends on the observable and the
centre-of-mass energy. For a theory defined up to ${\cal
O}(\alpha_s^2)$, the uncertainty is formally expected to be at
${\cal O}(\alpha_s^3)$. The matching of fixed order and resummed
calculations, however, may alter the scaling with $\alpha_s^3$.
The evolution of the uncertainty with $\alpha_s$ is shown in
Fig.~\ref{fig:asscal}. In this case the symmetric uncertainty is
analysed, {\it i.e.} the mean of the upward and downward
uncertainty. \EPSFIGURE{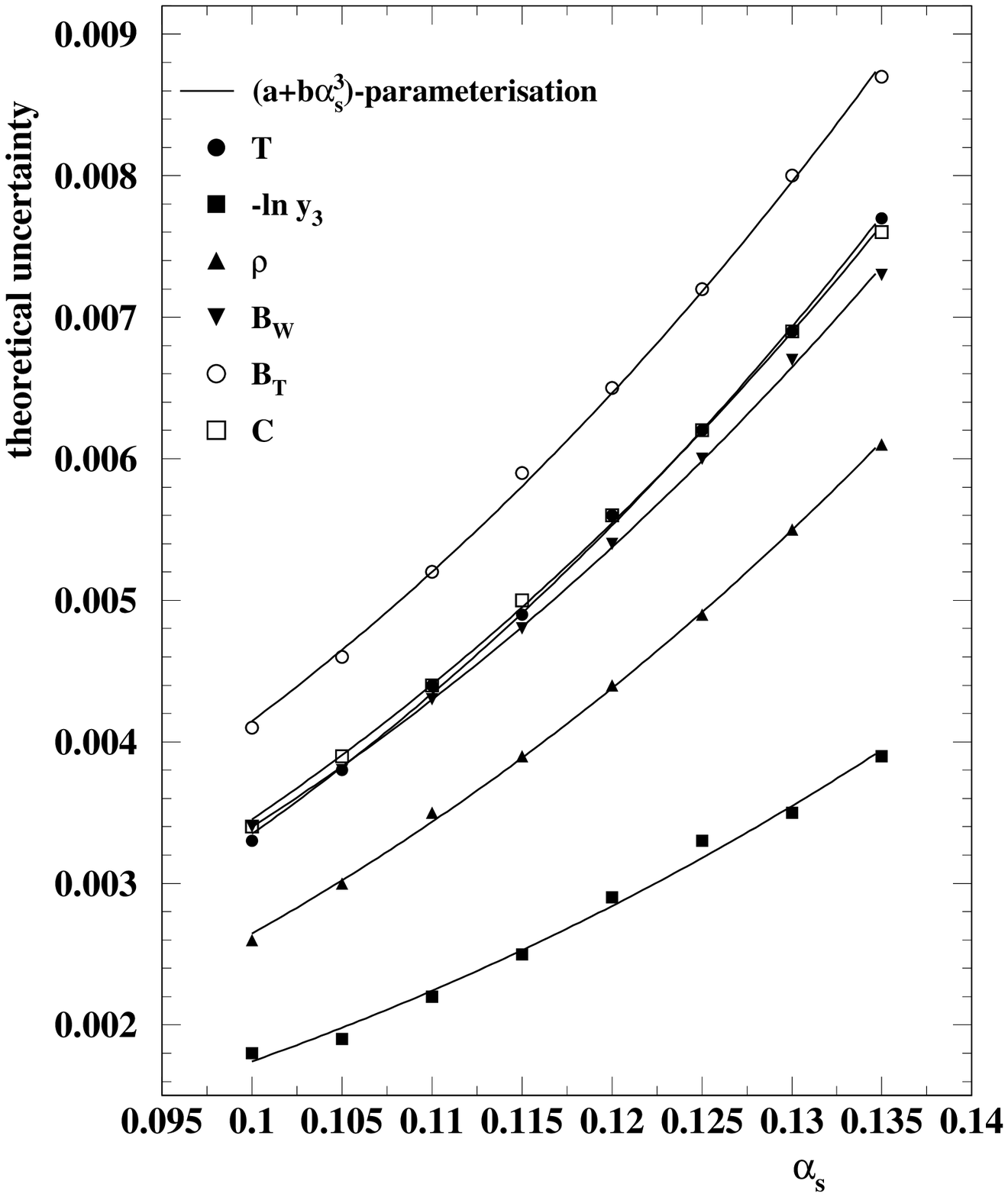,width=150mm} {Evolution
of the theoretical
  uncertainty with $\alpha_s$. The lines are the result of a
  parameterisation scaling with $\alpha_s^3$.\label{fig:asscal}}
A fit of the form $a + b \alpha_s^3$ gives a good description of the
dependence of the uncertainty on $\alpha_s$. The parameters $a$ and $b$
obtained with the fit are given in Table~\ref{tab:asscal} for each
observable. In general $a$ is found to be substantially smaller
than $b \alpha_s^3$, confirming that to a reasonable approximation
the simple $\alpha_s^3$ scaling discussed above is observed.
\TABULAR[ht!]{|l|c|c|} { \hline Observable & $a$ & $b$ \\ \hline $T$ &
  $3.6 \, 10^{-4}$ & $2.99$ \\ \hline $-\ln y_3$ & $2.4 \, 10^{-4}$ &
  $1.51$ \\ \hline $\rho$ & $2.7 \, 10^{-4}$ & $2.38$ \\ \hline $B_W$
  & $6.8 \, 10^{-4}$ & $2.71$ \\ \hline $C$ & $5.8 \, 10^{-4}$ &
  $2.88$ \\ \hline $B_T$ & $9.7 \, 10^{-4}$ & $3.18$ \\ \hline }
{\label{tab:asscal}Parameterisation of the uncertainty evolution with
  $\alpha_s$ according to a form $ a + b \alpha_s^3$.}
The two main components of the theoretical uncertainty originate from
the variations of $\xmu$ and $x_L$. The size of the uncertainties
depends crucially on the range of variation, nominally from $\frac{1}{2}$ to 2
for $\xmu$ and from $\frac{2}{3}$ to $\frac{3}{2}$ for $x_L$.  The
dependence on the range choice is studied by re-scaling the variation
range by a factor of $x_R$. Hence, the re-scaled range has an upper
limit of $x_{up}*x_R$ ($x_{up}=2$ resp. $\frac{3}{2}$ for $\xmu$ resp.
$x_L$) and a lower limit of $x_{down}/x_R$ ($x_{down}=\frac{1}{2}$
resp. $\frac{2}{3}$ for $\xmu$ resp. $x_L$). The effect of a range variation is
expected to be linear in the logarithm of the re-scaling factor $x_R$
and the results are
given as function of $\ln x_R$ for both the upper bound
(positive error) and lower bound (negative error).
\EPSFIGURE[ht!]{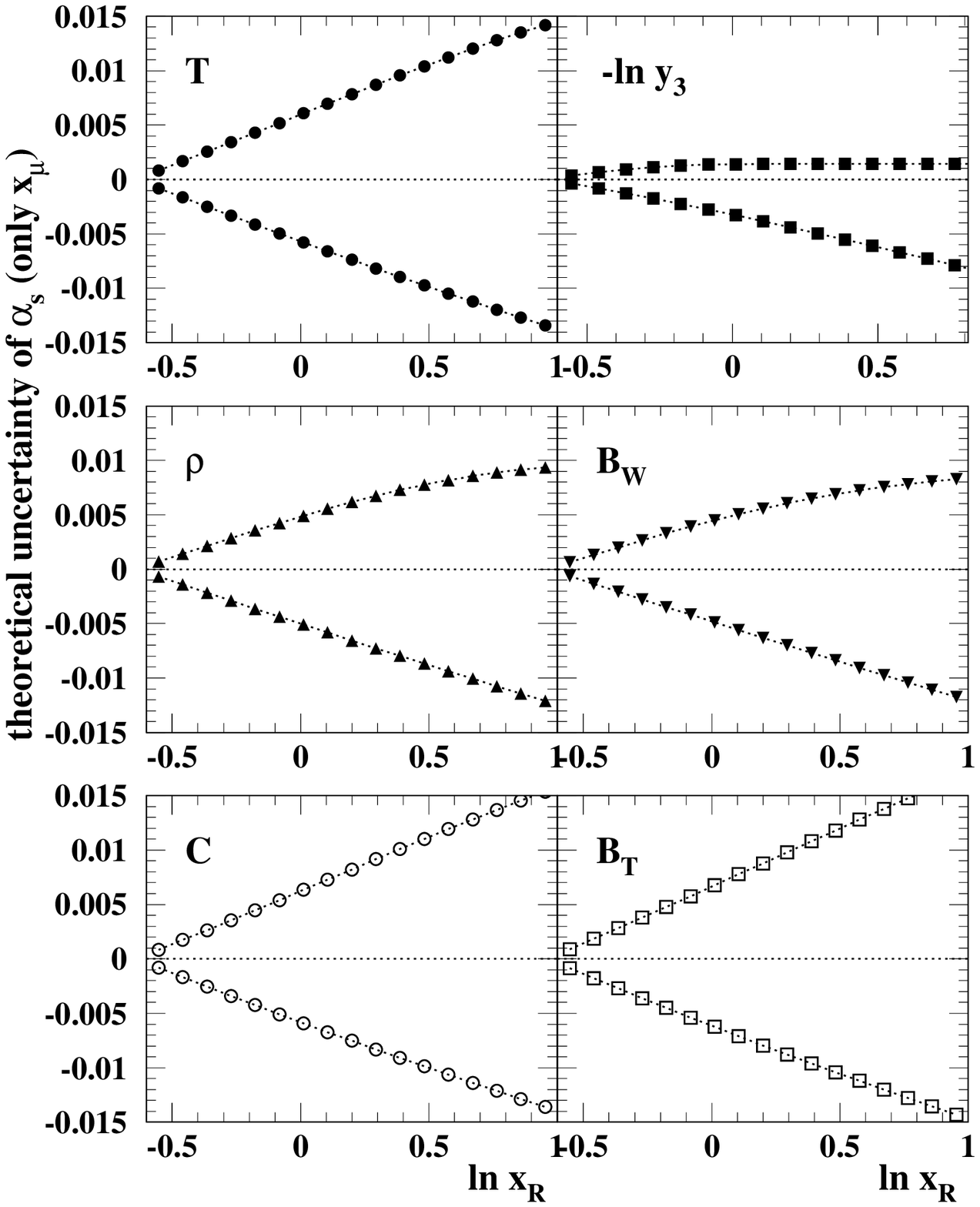,width=150mm} {The dependence of the positive and negative uncertainty
  related to the variation of the renormalisation scale $\xmu$. The width
of the variation range is modified such that
  the lower endpoint is rescaled by $1/x_R$ and the upper limit by
  $x_R$.\label{fig:xmu}}
\EPSFIGURE[ht!]{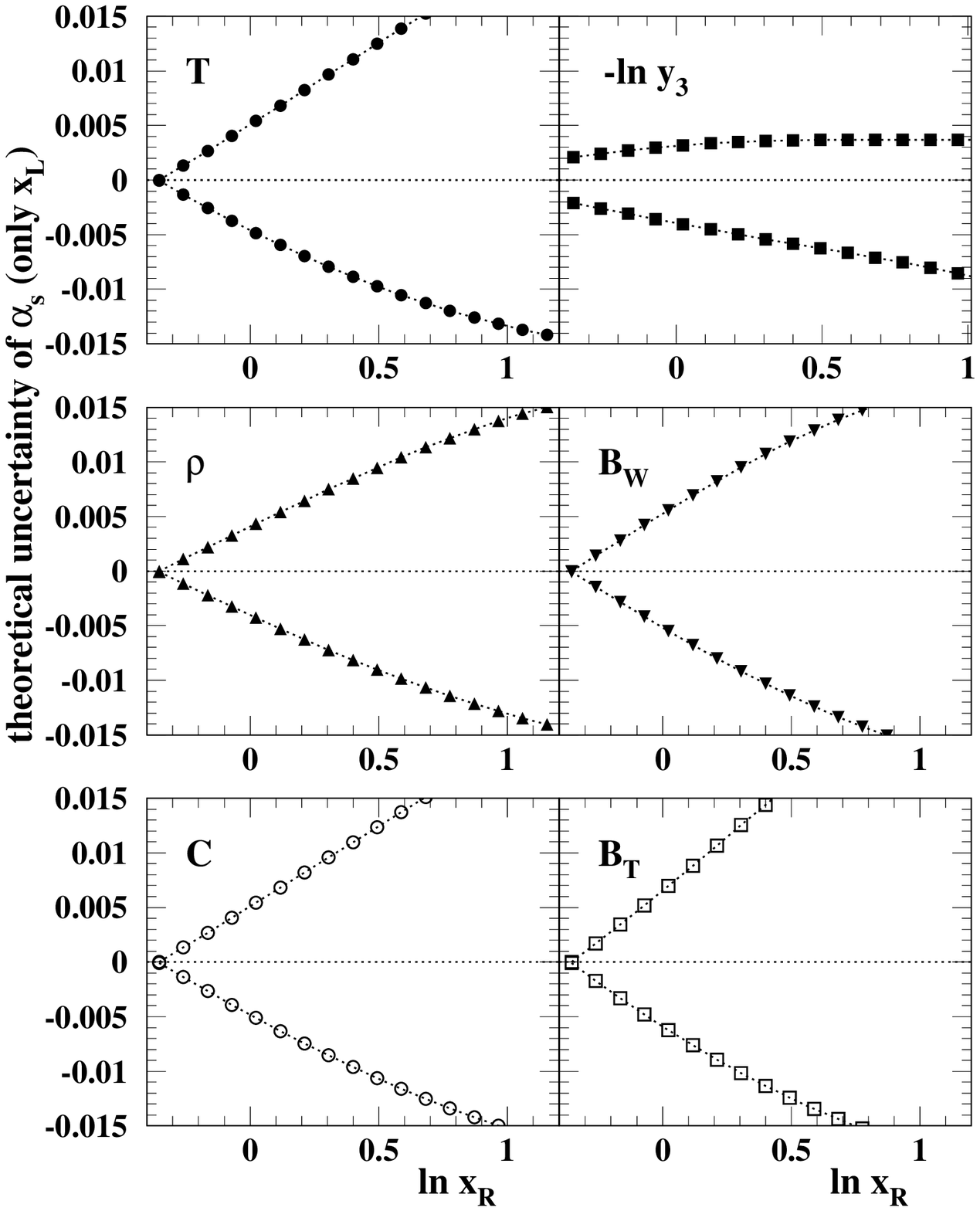,width=150mm}{The dependence of the positive and negative
uncertainty related to the variation of the logarithmic re-scaling factor $x_L$. The width
of the variation range is modified such that
  the lower endpoint is rescaled by $1/x_R$ and the upper
  limit by $x_R$.\label{fig:fitxl}}
The dependence of the $\xmu$ component of the uncertainty on the range
of variation is shown in Fig.~\ref{fig:xmu}. The size of uncertainty
increases rapidly with the width of the range,
the positive error flattens at large ranges for $-\ln y_3$, $\rho$ and $B_W$.
The shape of the dependence is similar for all variables. The case of the $x_L$ component
is shown in Fig.~\ref{fig:fitxl}. Here the rise at small variation
ranges is even steeper than for $\xmu$. For the variable $-\ln y_3$
the evolution is significantly flatter than for all other variables, a
similar effect would have been observed with a quadratic re-scaling of
the variation range.

Finally, the dependence of the uncertainty on the fit range experimentally used
to extract $\alpha_s$ from the distribution is studied. For this purpose the uncertainty band method including
all uncertainty sources is applied to each bin of the distribution, as shown in Fig.~\ref{fig:fitra}.
In the central part of the distribution the uncertainty is reasonably constant and stable
with respect to variations of one or two bins at each end. Approaching the peak of the distributions
or the extreme three-jet region, however, induces rapidly growing uncertainties. This dependence
should be kept in mind when selecting a fit range.
\EPSFIGURE[hb!]{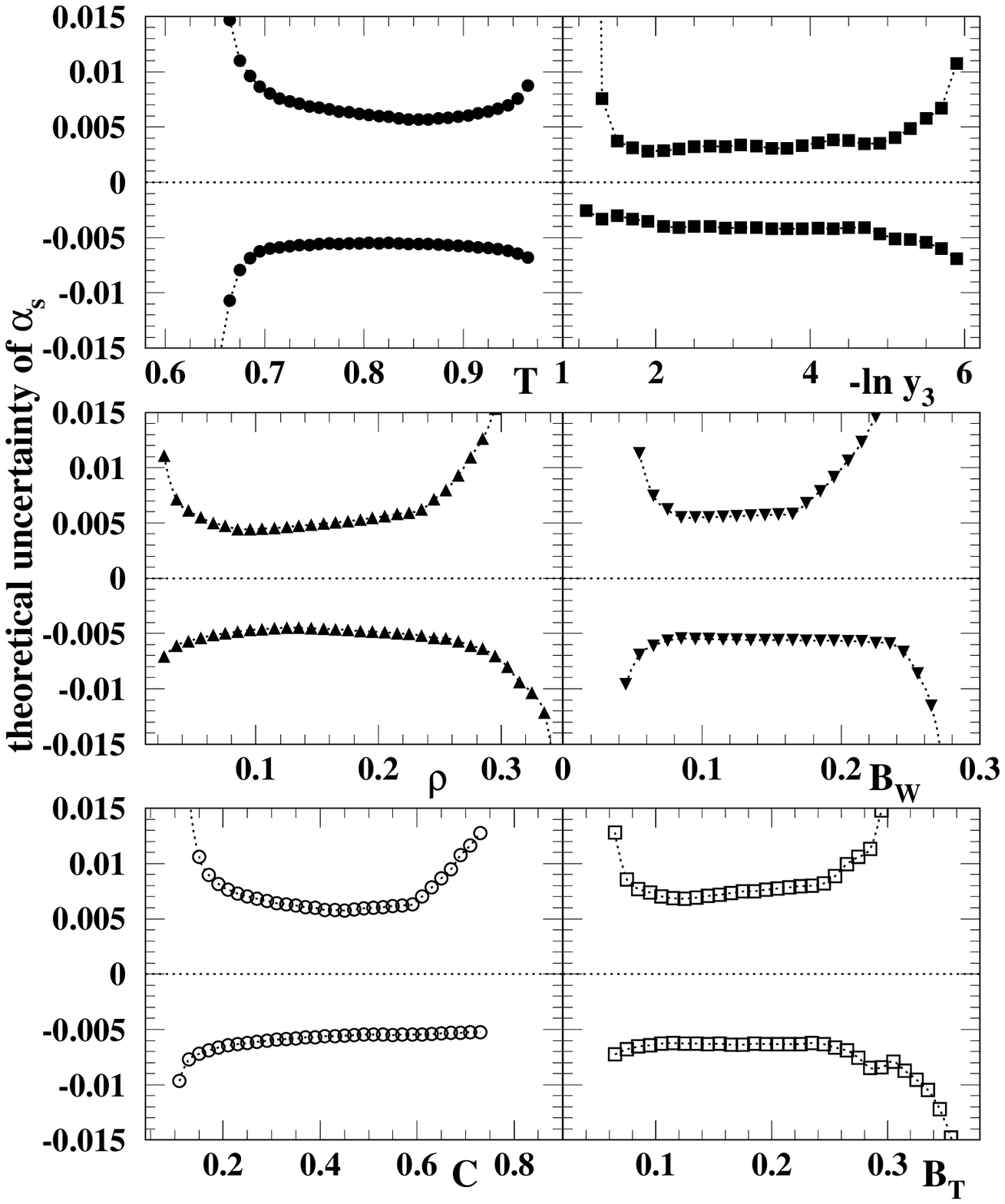,width=150mm}{The dependence of the positive and negative uncertainty
  on the value of the observable. All components of the uncertainty are included.
  \label{fig:fitra}}

\clearpage

The theory uncertainty studies described here are all carried out
on purely perturbative predictions. One could instead envisage
carrying out these studies after correction to hadron-level.
Whether this actually makes a difference or not depends on the
details of how the hadronisation is included, for example whether
as a bin-by-bin multiplicative factor, a simple shift of the
distribution, or a transfer matrix. However, since hadronisation
corrections will have similar effects both on the various
alternate theory curves and on the reference prediction, the net
impact of hadronisation is expected to be rather small and hence
it is simplest and most transparent to carry out the analysis at a
purely perturbative level

\section{Conclusion}\label{sec:conc}
In this paper, a new method is presented for the
  assessment of theoretical uncertainties in $\alpha_s$.  This method
  evaluates the systematic uncertainty of the parameter $\alpha_s$
  from the uncertainty of the prediction for the distributions from which
  it is extracted.
After a comprehensive review of theoretical predictions for measurements
of $\alpha_s$ from event-shape distributions in $\epem$ annihilation,
the uncertainties of such predictions are estimated with an uncertainty
band method which incorporates several variations of the theory differing
in subleading terms and includes a new test for re-scaling
the resummed logarithmic variables. As the uncertainty band method can compute
these perturbative uncertainties of $\alpha_s$ independently of a measured
event-shape distribution, it is especially suited for an unbiased combination of several
observables or experiments.

The recommended method for computing the central result is to use
the modified Log(R) matching scheme with $\xmu=x_L=1$, $p=1$ and
values for $\ymax$ given in Table~\ref{tab:ymax}. The assessment
of the perturbative uncertainty consists of a variation of the
renormalisation scale $\xmu$ from 0.5 to 2, of the logarithmic
re-scaling factor $x_L$ from $\frac{2}{3}$ to $\frac{3}{2}$, of a
replacement of the modified Log(R) matching scheme by the modified
R matching scheme, of the degree of modification $p=1$ by $p=2$
and of the kinematic constraint $\ymax$ by its alternative
$y^\prime_{\rm max}$ given in Table~\ref{tab:ymax}. The different
uncertainties should be combined with the uncertainty band method.
Proposals are made for more conservative uncertainty estimates;
these are a larger range for the re-scaling factor $|\ln x_L|<0.6$
and an alternative operating mode of the uncertainty band method.

To combine several measurements from different observables or
experiments the perturbative uncertainties of the individual
measurements should be re-estimated with the above method using a
common input value for $\alpha_s(\mz)$ in the uncertainty band
method. This input value should of course be consistent with the
final result obtained iteratively.

Following these instructions even existing measurements of $\alpha_s$ can be
equipped with an up-to-date estimation of their perturbative uncertainty and
then used in a consistent combination.

\section*{Acknowledgements}

We wish to thank Stefano Catani for helpful suggestions and G\"unther Dissertori, Einan Gardi,
Klaus Hamacher and Oliver Passion for useful conversation on this subject.

\listoftables       
\listoffigures      

\end{document}